\documentclass[5p,twocolumn]{elsarticle}

\usepackage{graphicx}
\usepackage{dcolumn}
\usepackage{pifont}
\usepackage{bm}
\usepackage{multirow}
\usepackage{amsmath}
\usepackage{float}
\usepackage{txfonts}
\usepackage[version=3]{mhchem}

\usepackage[usenames,dvipsnames]{xcolor}
\definecolor{blue}{RGB}{0,0,225}
\definecolor{cream}{RGB}{222,217,201}
\definecolor{red}{RGB}{225,0,0}

\journal{arXiv}

\begin{document}
\title{Computational Prediction of Structural, Electronic, Optical Properties and Phase Stability of Double Perovskites K$_2$SnX$_6$ (X = I, Br, Cl)}

\author[kimuniv-m]{Un-Gi Jong}
\author[kimuniv-m]{Chol-Jun Yu\corref{cor}}
\cortext[cor]{Corresponding author}
\ead{cj.yu@ryongnamsan.edu.kp}
\author[kimuniv-m]{Yun-Hyok Kye}

\address[kimuniv-m]{Chair of Computational Materials Design, Faculty of Materials Science, Kim Il Sung University, Ryongnam-Dong, Taesong District, Pyongyang, Democratic People's Republic of Korea}

\begin{abstract}
Vacancy-ordered double perovskites \ce{K2SnX6} (X = I, Br, Cl) attract significant research interest due to their potential application as light-absorbing materials in perovskite solar cells.
However, a deep insight into their material properties at the atomic scale is yet scarce.
Here we present a systematic investigation on their structural, electronic, optical properties and phase stabilities in cubic, tetragonal, and monoclinic phases based on density functional theory calculations.
Quantitatively reliable prediction of lattice constants, band gaps, effective masses of charge carriers, exciton binding energies is provided in comparison with the available experimental data, revealing the increasing tendency of band gap and exciton binding energy as lowering the crystallographic symmetry from cubic to monoclinic and going from I to Cl.
We highlight that cubic \ce{K2SnBr6} and monoclinic \ce{K2SnI6} are suitable for the application as a light-absorber for solar cell devices due to their proper band gaps of 1.65 and 1.16 eV and low exciton binding energies of 59.4 and 15.3 meV, respectively.
The constant-volume Helmholtz free energies are determined through phonon calculations, giving a prediction of their phase transition temperatures as 449, 433 and 281 K for cubic-tetragonal and 345, 301 and 210 K for tetragonal-monoclinic transitions for X = I, Br and Cl.
Our calculations provide an understanding of material properties of vacancy-ordered double perovskite \ce{K2SnX6}, helping to devise a low-cost and high performance perovskite solar cell.
\end{abstract}

\maketitle

\section{Introduction}
During the past decade, organic-inorganic hybrid halide perovskites revolutionized the field of photovoltaics with a sharply rapid progress in power conversion efficiency (PCE) and remarkably lower fabrication cost, compared with the traditional silicon solar cells~\cite{Kojima, Zhou14, Jeon, WSYang}.
The prototype perovskite solar cells (PSCs) adopt usually methylammonium lead iodide (\ce{CH3NH3PbI3} or \ce{MAPI3}) as a light absorbing material, being composed of earth abundant elements and exhibiting unique properties beneficial to the solar cell applications~\cite{Yamada, Xing, Frost14, Philippe, Wolf, Burschka, Baikie}.
Using this kind of hybrid halide perovskites, PCE reached the certified record of 22.7\%~\cite{JFeng} in 2018, being high enough for solar cell application, but they hold also fatal problems of instability to humidity, temperature and light and moreover toxicity of lead, which still hinder the comperciallization of PSCs~\cite{Noh, Luo15, Kye18jpcl, Philippe, Yang, Berhe}.

These challenges could be addressed partially by utilizing solid solutions through mixing Br or Cl with I anion~\cite{Aharon, Noh, Sadhanala, Jong16prb} and another organic moiety formamidium (FA) or even inorganic Cs and Rb with MA cation.
Noh et al.~\cite{Noh} demonstrated that the stability of mixed-halide perovskite \ce{MAPb(I_{1-x}Br_x)3} was significantly improved with relatively high PCEs as the mixing ratio $x<0.2$, and similar effect was found when partially replacing I with Cl in \ce{MAPb(I_{1-x}Cl_x)3}~\cite{Tripathi, Ng, jong2017jps}.
Alternatively, double or triple mixed-cation perovskites have been found to substantially improve the efficiency and phase stability~\cite{Niu, ZLi, Rehman, Lee, Saliba, Duong, Zhang}.
For instance, Niu et al.~\cite{Niu} fabricated PSCs with a composition of \ce{Cs$_x$MA$_{1-x}$PbI3}, reporting that introducing small amount of Cs into MA ($x\sim0.09$) was beneficial not only to better stability but also to higher efficiency of solar cells.
By using a mixture of tripe Cs/MA/FA cation, Saliba et al.~\cite{Saliba} further achieved higher peak efficiency of 21.1\% and 18\% under the operational condition after 250 hours.
On the other hand, mixed-cation and mixed-halide perovskites like \ce{Cs$_y$FA$_{1-y}$Pb(Br$_x$I$_{1-x}$)3}~\cite{Rehman} was also examined to tune the phase stability, photo stability and optoelectronic properties by carefully changing the chemical compositions of cation $y$ and halide anion $x$.

\begin{table*}[!th]
\small
\caption{\label{tab_lattice} Goldschmidt's tolerance factor ($t_G$), octahedral factor ($t_o$), radius ratio ($t_r$), lattice Constant ($a, b, c$) and angle ($\beta$) of \ce{K2SnX6} (X = I, Br, Cl) in cubic, tetragonal, and monoclinic phases, calculated by using PBE functional.}
\begin{tabular}{lllllll}
\hline
material & phase & $t_G$ & $t_o$ & $t_r$ & cal. & exp.~\cite{Boysen, Higashi} \\
\hline
\ce{K2SnCl6} & cub. & 0.88 & 0.39 & 0.89 & $a=10.02$~(\AA)& $a=9.99$~(\AA)\\
& tet. &      &      &      & $a=7.09,~b=10.01$~(\AA)& $a=7.06,~b=9.98$~(\AA)\\
& mon. &      &      &      & $a=7.07,~b=7.05,~c=10.03$~(\AA)& $a=7.02,~b=7.01,~c=9.99$~(\AA)\\
&      &      &      &      & $\beta=90.11$~(deg) & $\beta=90.13$~(deg)\\      
\hline
\ce{K2SnBr6} & cub. & 0.87 & 0.36 & 0.85 & $a=10.51$~(\AA)& $a=10.48$~(\AA)\\
& tet. &      &      &      & $a=7.50,~b=10.67$~(\AA)& $-$\\
& mon. &      &      &      & $a=7.45,~b=7.47,~c=10.68$~(\AA)& $a=7.43,~b=7.44,~c=10.62$~(\AA)\\
&      &      &      &      & $\beta=90.17$~(deg) & $\beta=90.18$~(deg)\\      
\hline
\ce{K2SnI6} & cub.  & 0.85 & 0.32 & 0.79 & $a=11.66$~(\AA)& $-$\\
& tet.  &      &      &      & $a=8.25,~b=11.76$~(\AA)& $-$\\
& mon.  &      &      &      & $a=8.29,~b=8.32,~c=11.69$~(\AA)& $-$\\
&       &      &      &      & $\beta=90.25$~(deg) & $-$\\
\hline
\end{tabular}
\end{table*}

In parallel with these attempts in the hybrid organic-inorganic halide perovskites, fully replacing the organic cation with the inorganic Cs or Rb cation, resulting in the all-inorganic perovskites, has been regarded as a favorable way for improving the stability because of lower sensitivity of the inorganic cations to moisture~\cite{Hoffman, Kulbak, Eperon, Heidrich, Bekenstein, Swarnkar, Akkerman}.
In fact, many papers reported that PSCs adopting the inorganic cesium lead iodide perovskite (\ce{CsPbI3}) exhibited a high PCE comparable to the hybrid PSCs and significantly enhanced device stability~\cite{Eperon, Hoffman, Swarnkar}.
However, it is challenging for \ce{CsPbI3} to form the photoactive black phase with cubic lattice ($\alpha$-\ce{CsPbI3}) at room temperature~\cite{Eperon, Moller}, and moreover it still contains toxic lead element.
Accordingly, \ce{CsSnI3} has been suggested to be used as a potential candidate of non-hygroscopic and non-toxic halide perovskite.
Several groups have synthesized the photoactive cubic \ce{CsSnI3} stable at room temperature, but the performance of \ce{CsSnI3}-based PSCs have shown too lower PCE of up to 2\%~\cite{Chen12, Kumar14, Sabba15}.
Theoretical and experimental investigations suggested that such a poor efficiency comes from the problem of ready oxidation of tin cation from Sn$^{2+}$ to Sn$^{4+}$ state being subject to the deterioration of optoelectronic properties of \ce{CsSnI3}~\cite{XLi, Stoumpos, Jung17cm}.

Structural diversity of perovskite material can open an outlet for avoiding such severe oxidation, i.e., vacancy-ordered double perovskite \ce{Cs2SnI6}, which is obtained by removing every other Sn cation from the fully occupied conventional perovskite \ce{CsSnI3}.
It was reported that \ce{Cs2SnI6} adopts the cubic phase at room temperature with a direct band gap of about 1.3 eV, strong visible light absorption coefficients, long carrier mobilities, and air and moisture stability, all of which are advantageous for solar cell application~\cite{Maughan16jacs, Maughan18cm1, Maughan18cm2, LeeB, Saparov}.
Despite these merits, there unfortunately exists an obstacle to a wide application of \ce{Cs2SnI6} for large-scale and low-cost PSCs due to the small amount of cesium in the earth's crust.
In fact, cesium occupies only 0.00019\% of weight of the earth's crust, being considered as the 50-th common element in the periodic table.
This motivated to use potassium instead, the same group element as cesium, which is the 7-th common element with an amount of 2.6\% of weight in the earth's crust, being 10 000 times larger than the amount of cesium~\cite{csk}.
Therefore, replacing cesium with potassium satisfies the criterion for realizing the stable, environment-friendly, large-scale and low-cost all-inorganic PSCs with the utilization of the earth-abundant elements.
Back to the late 1970s, only a few investigations had focused on the structural properties of \ce{K2SnBr6} and \ce{K2SnCl6}~\cite{Boysen, Higashi}, and thus, a comprehensive research on the potassium tin halide vacancy-ordered double perovskites \ce{K2SnX6} (X=I, Br, Cl) is indispensible for their photovoltaic applications.

In the present work, we perform density functional theory (DFT) calculations to explore structural, electronic, optical properties and phase stability of vacancy-ordered double perovskite \ce{K2SnX6} (X=I, Br, Cl), aiming at finding a possibility of their solar cell application.
Keeping in mind that vacancy-ordered double perovskites generally undergoe a series of phase transition from monoclinic to tetragonal, and to cubic phase upon increasing temperature, we begin by optimizing the crystal structures of \ce{K2SnX6} in the cubic (space group $Fm$\={3}$m$), tetragonal ($P4/mnc$) and monoclinic ($P2_1/n$) phases.
Using these optimized structural parameters, we calculate the electronic and optical properties including electronic energy bands with density of states (DOS), effective masses of charge carriers, dielectric constants, exciton binding energies and light-absorption coefficients, providing a systematic comparison of these properties as changing the halogen atom and crystalline symmetry.
Finally, by using density functional perturbation theory (DFPT) method, we determine the phonon dispersion curves with phonon DOS, and estimate the phase transition temperatures from cubic to tetragonal and to monoclinic phases based on the obtained constant-volume Helmholtz free energy.

%
\begin{figure*}[!t]
\small
\begin{center}
\begin{tabular}{ccc}
\includegraphics[clip=true,scale=0.32]{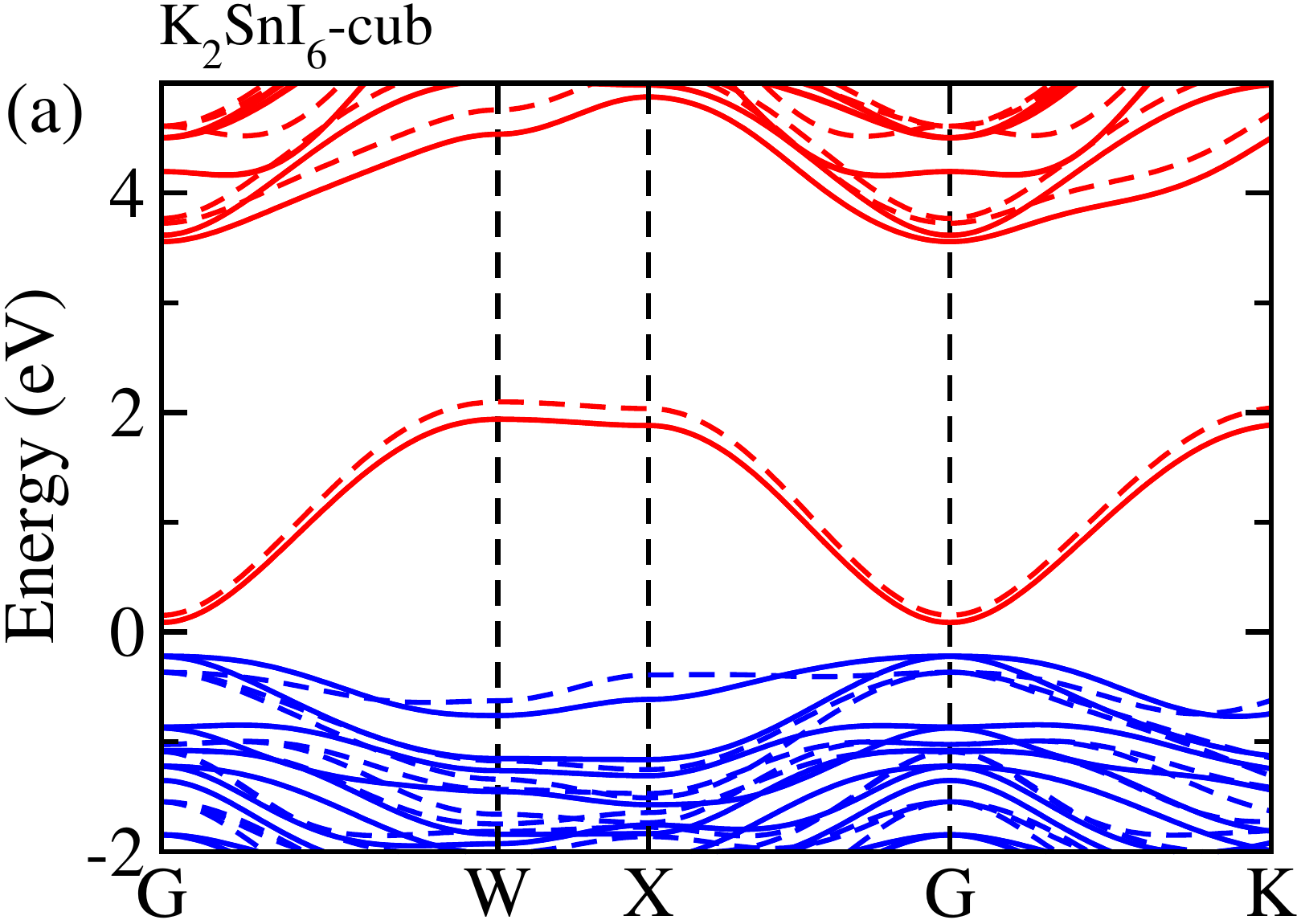} &
\includegraphics[clip=true,scale=0.32]{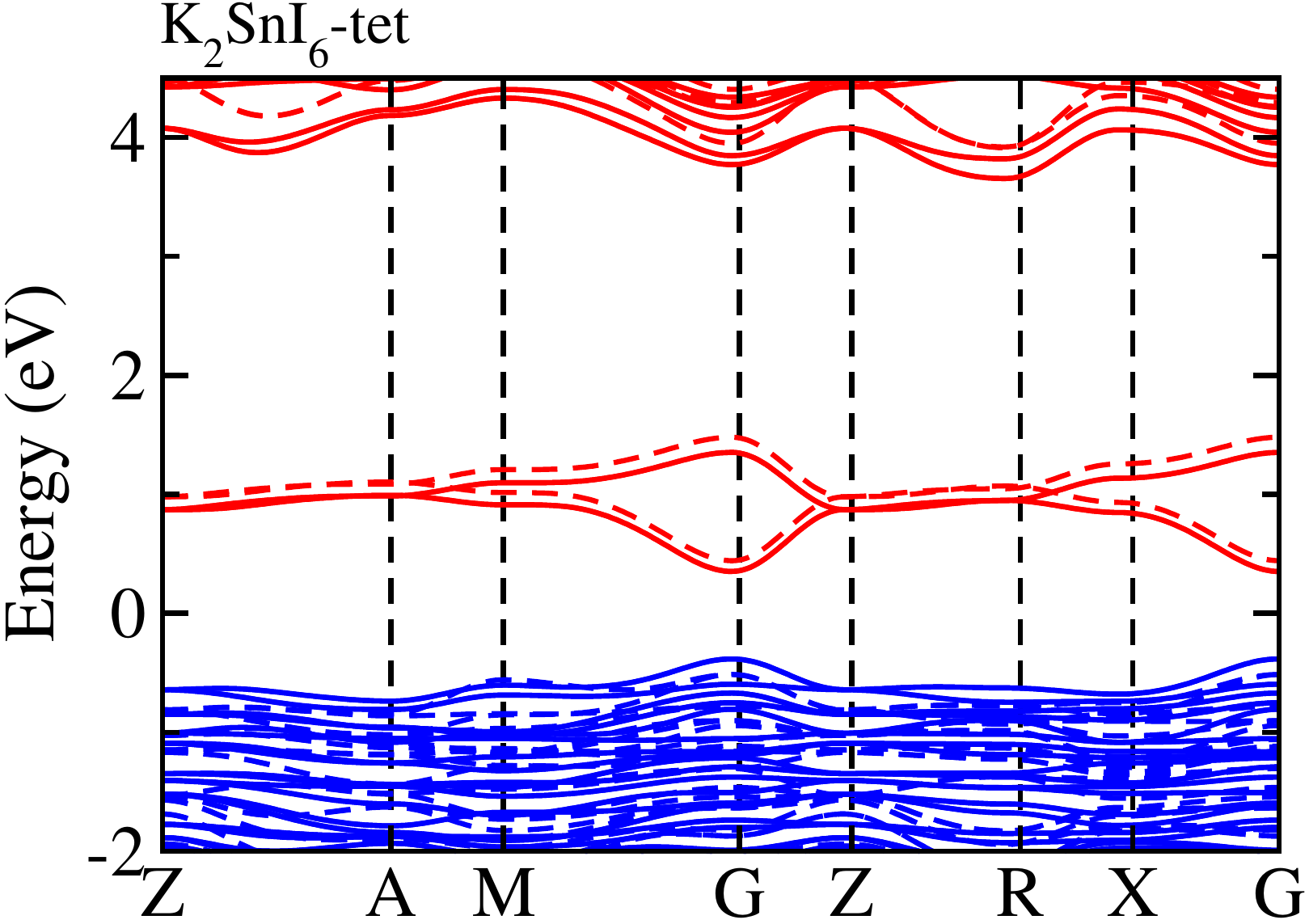} &
\includegraphics[clip=true,scale=0.32]{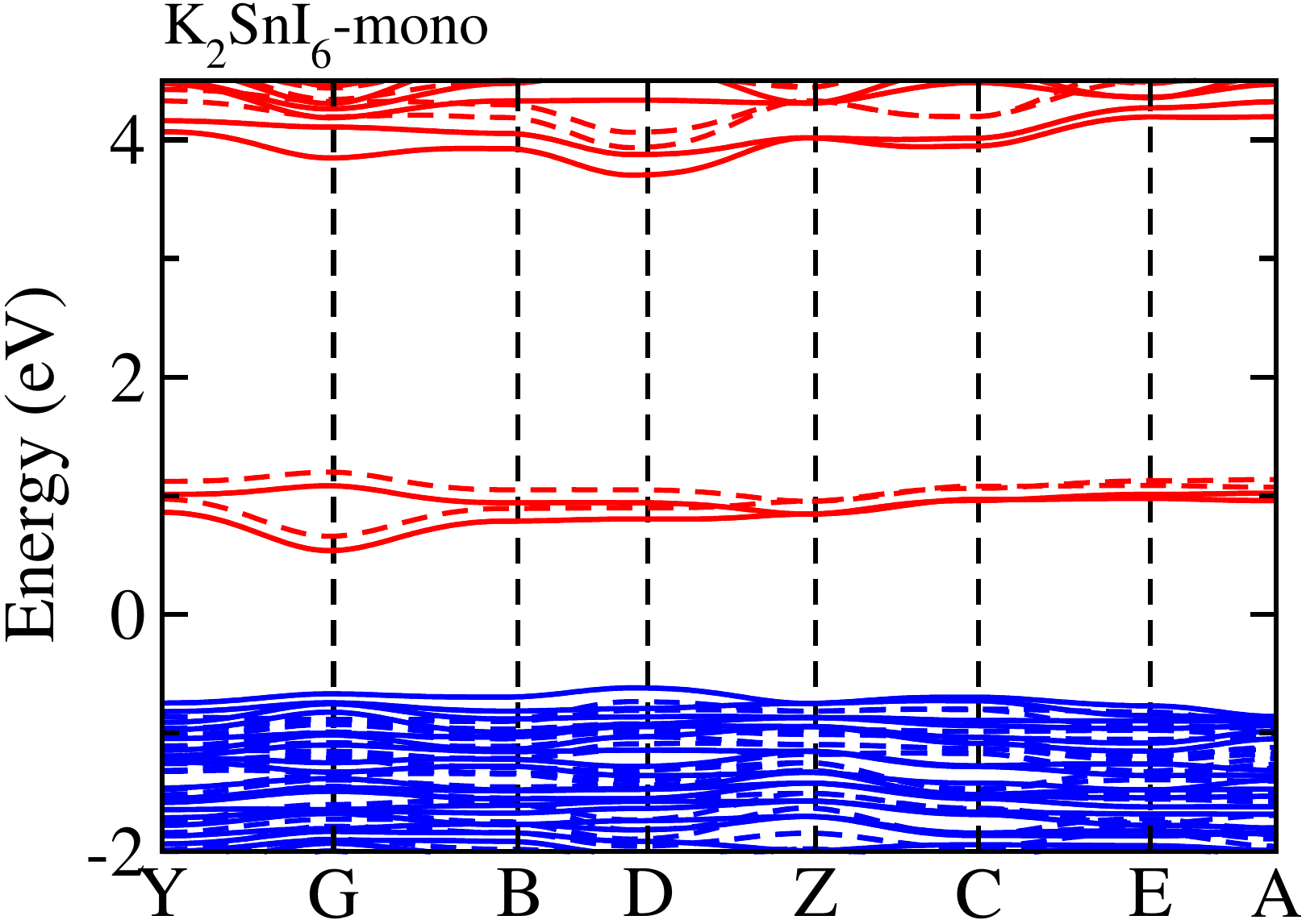} \\
\includegraphics[clip=true,scale=0.32]{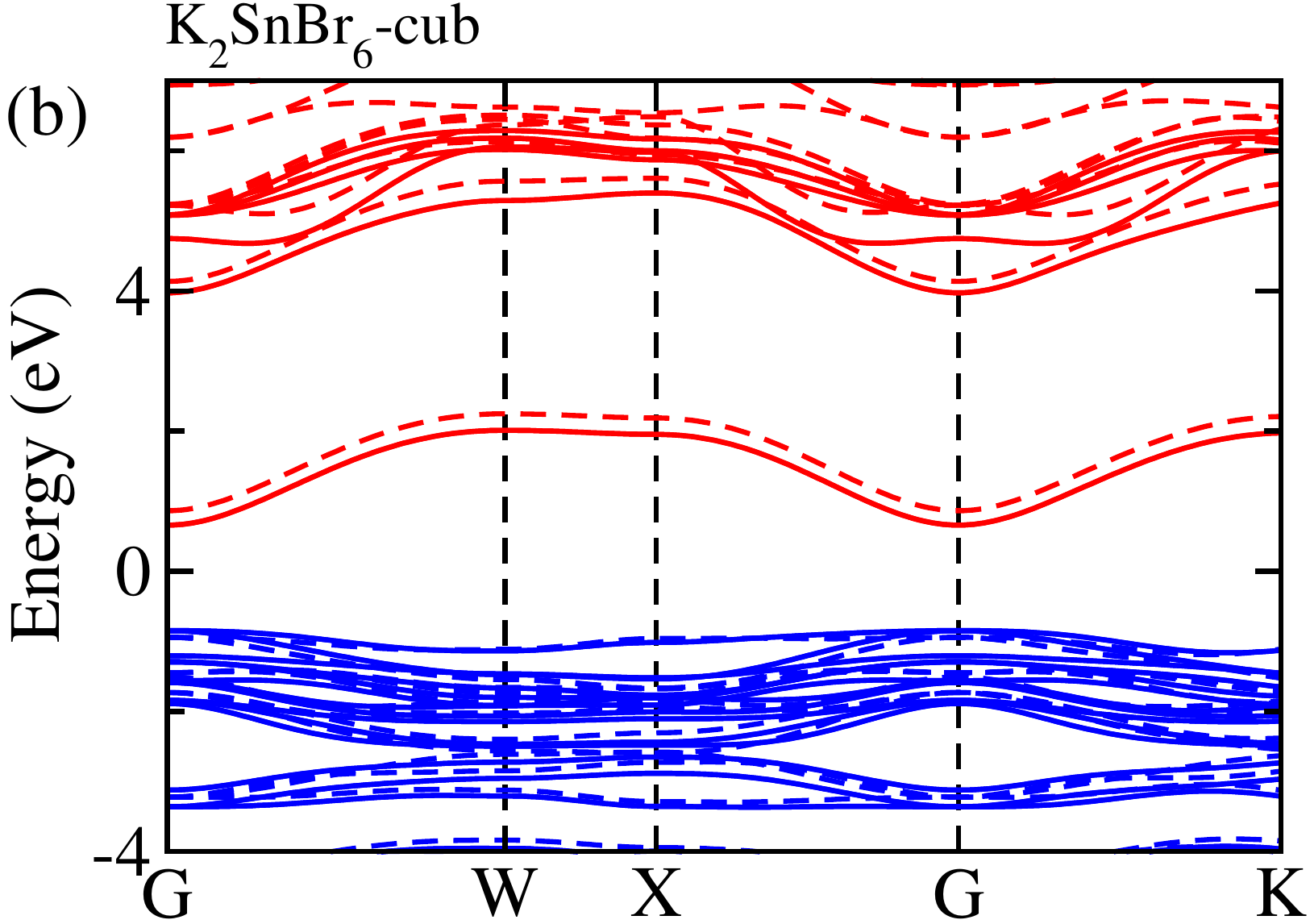} &
\includegraphics[clip=true,scale=0.32]{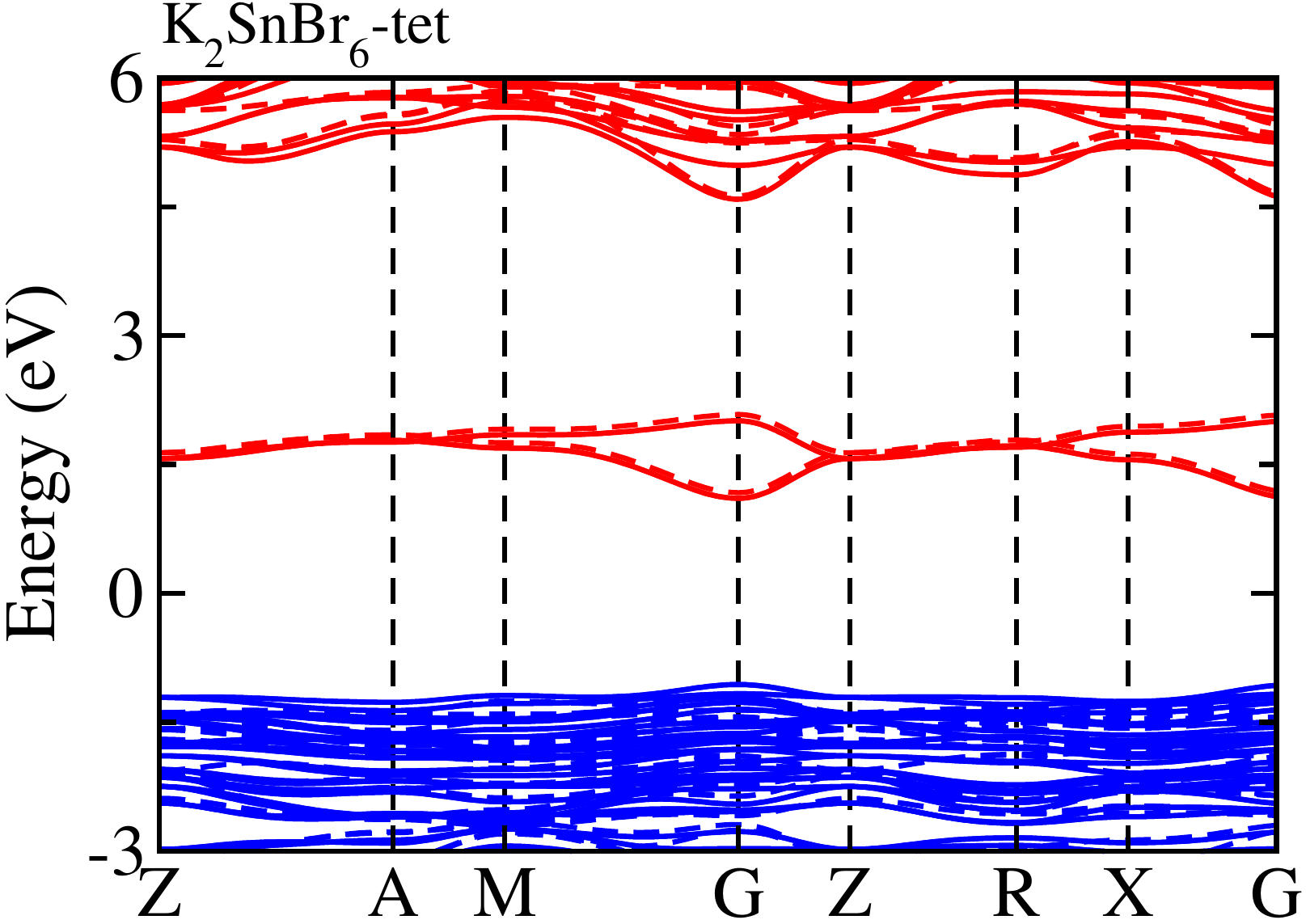} &
~\includegraphics[clip=true,scale=0.32]{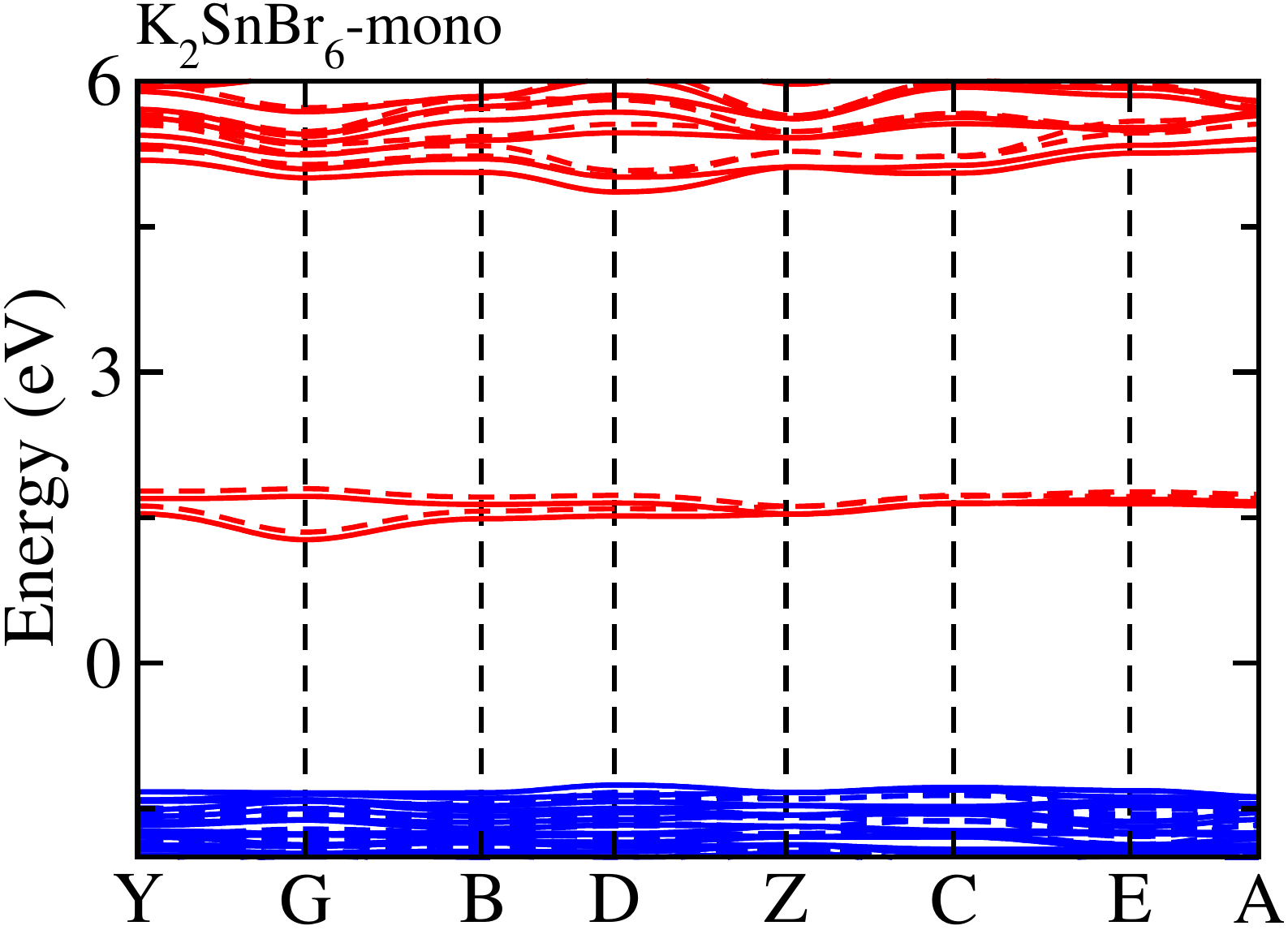} \\
\includegraphics[clip=true,scale=0.32]{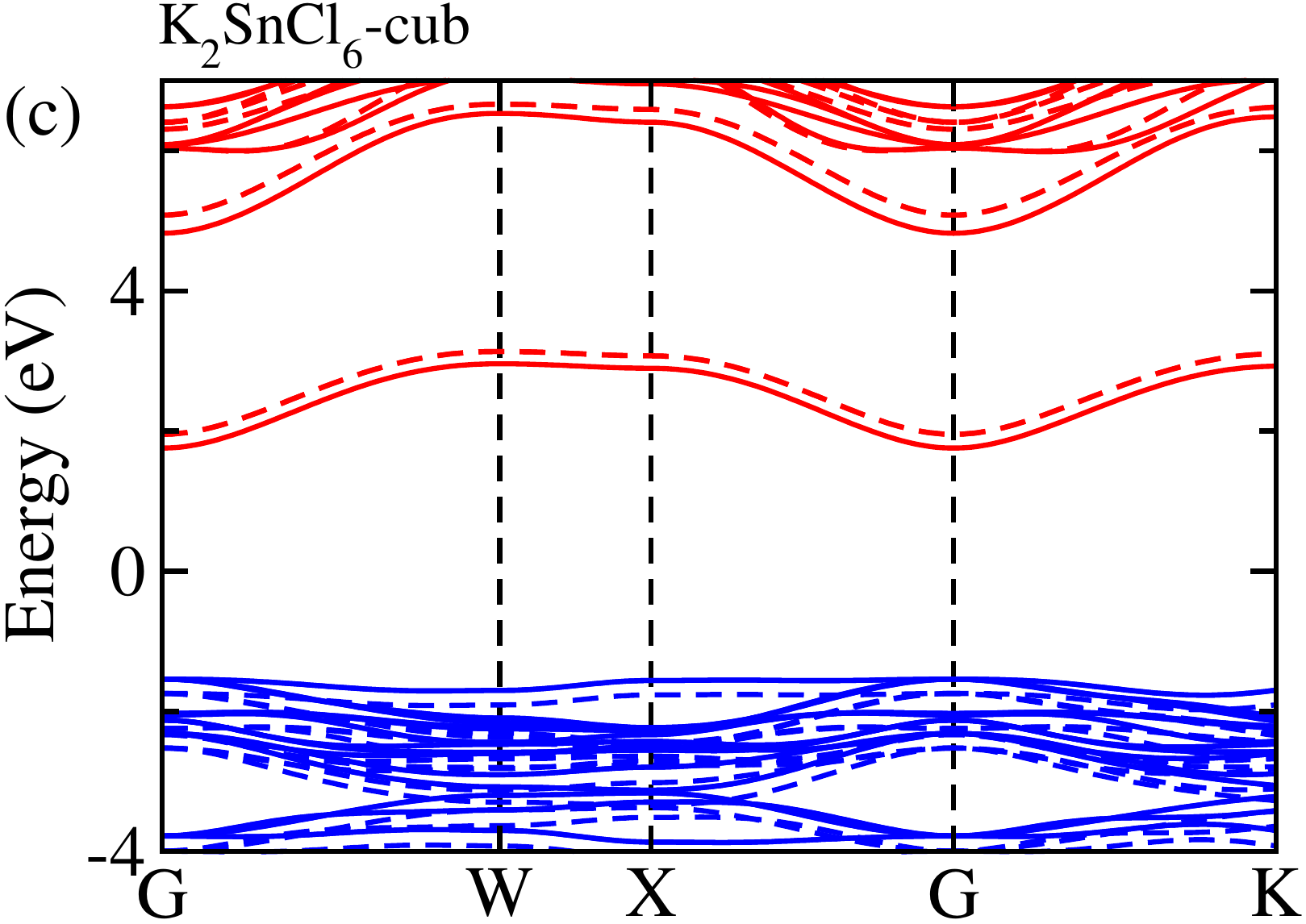} &
\includegraphics[clip=true,scale=0.32]{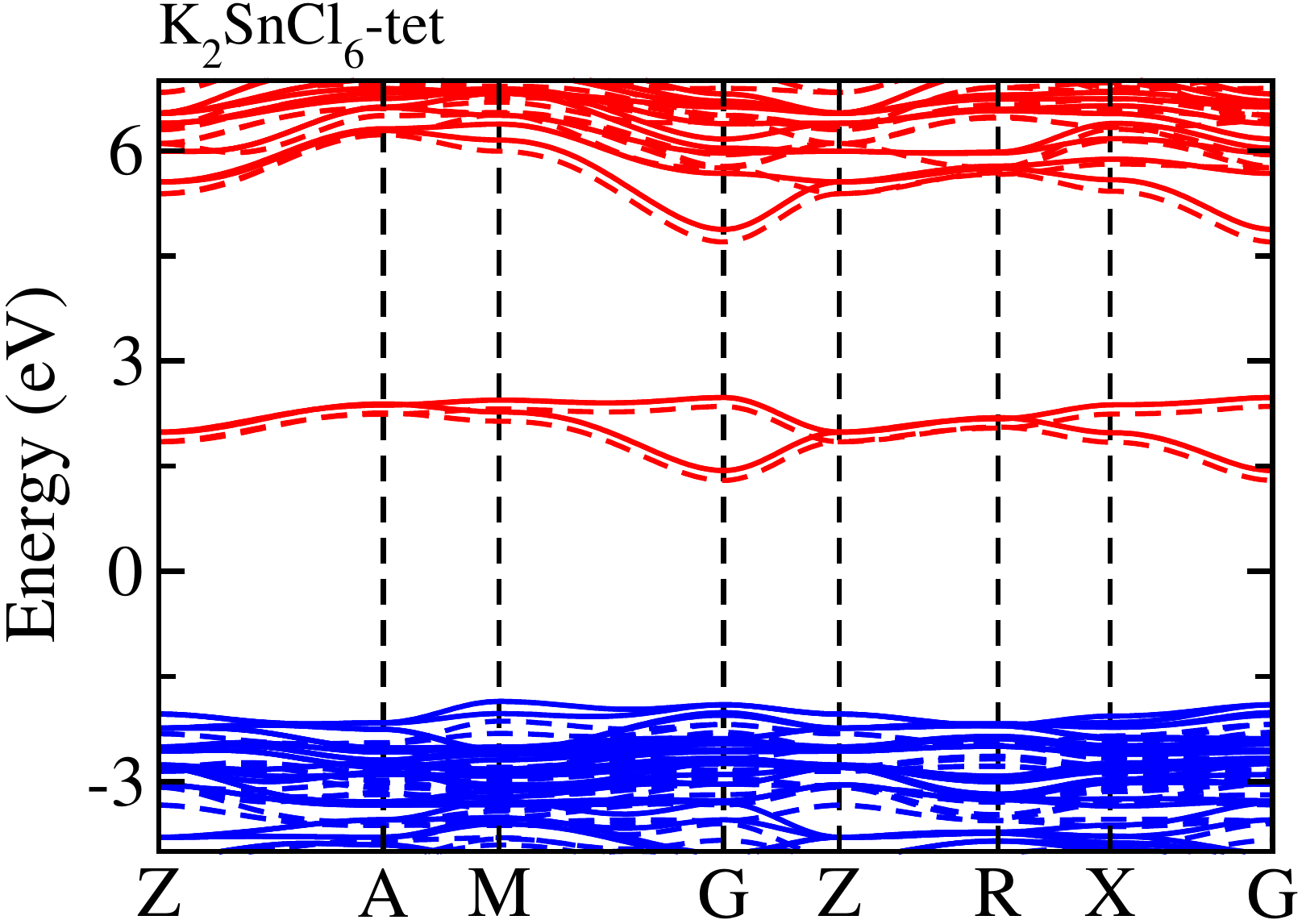} &
\includegraphics[clip=true,scale=0.32]{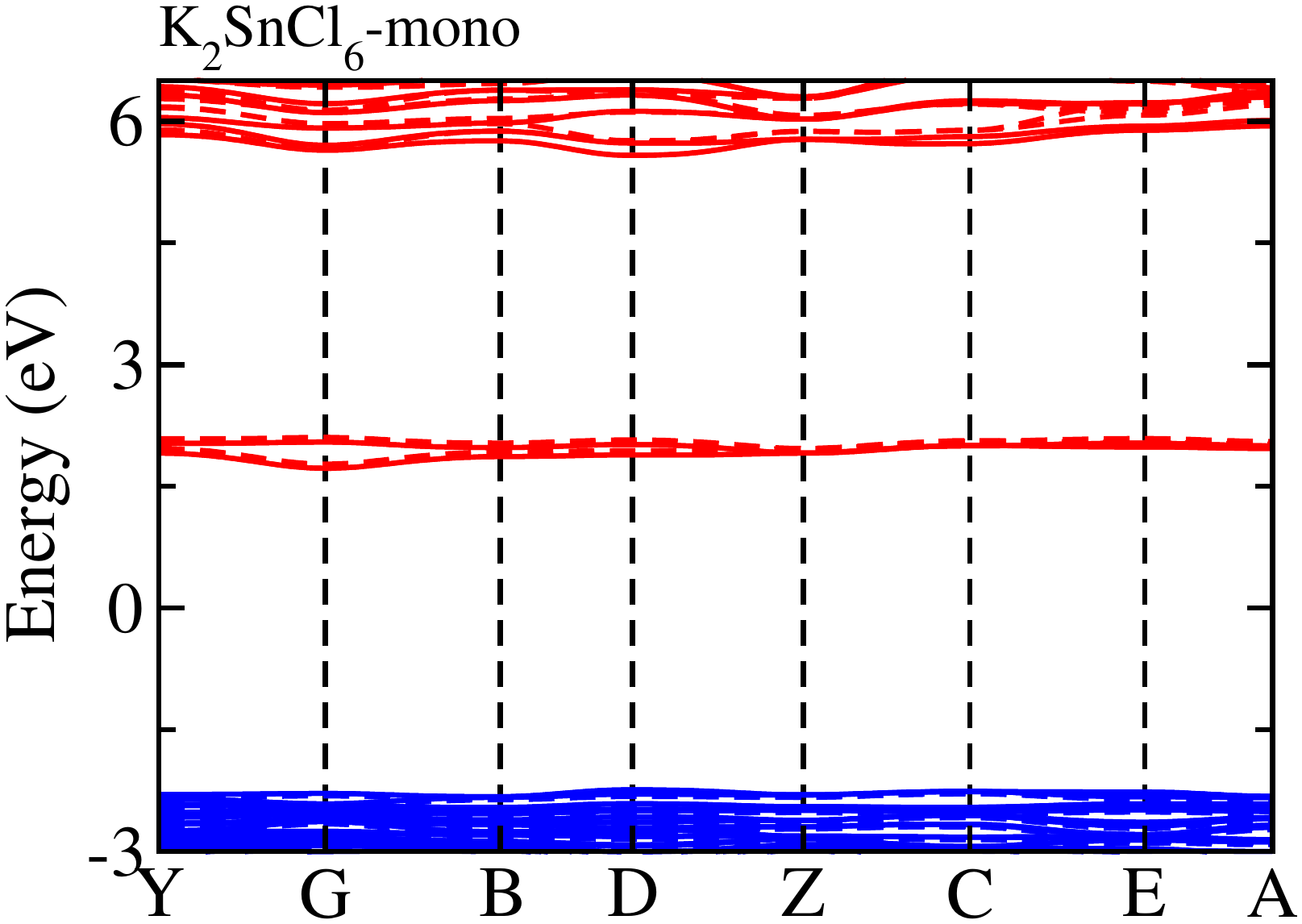} \\
\end{tabular} 
\end{center}
\caption{\label{fig_band}Electronic band structures of (a) \ce{K2SnI6}, (b) \ce{K2SnBr6} and (c) \ce{K2SnCl6} in cubic (left), tetragonal (middle) and monoclinic (right) phases, calculated by HSE06 hybrid functional with (solid line) and without (dashed line) spin-orbit coupling. Blue and red colors indicate valence and conduction bands.}
\end{figure*}

\section{Results and Discussion}
\subsection{Crystal structures}
Vacancy-ordered double perovskite is generally regarded to successively adopt cubic, tetragonal, and monoclinic phases upon decreasing temperature.
In fact, \ce{K2SnCl6} and \ce{K2SnBr6} were experimentally found to stabilize in the cubic, tetragonal, and monoclinic phases at different temperatures~\cite{Boysen, Higashi}.
Thus, as a first step, we performed optimizations of \ce{K2SnX6} (X = I, Br, Cl) crystal structures with the cubic ($Fm$\={3}$m$), tetragonal ($P4/mnc$) and monoclinic ($P2_1/n$) phases, as shown in Figure S1.
Table~\ref{tab_lattice} presents the result of the Goldschmidt's tolerance factor $t_G$, octahedral factor $t_o$, radius ratio $t_r$, lattice constant ($a, b, c$) and angle ($\beta$), calculated by using the PBE-GGA functional in comparison with the available experimental data~\cite{Boysen, Higashi} for \ce{K2SnX6} (X = I, Br, Cl).

As in the conventional perovskite \ce{ABX3}, we assessed formability of perovskite structure in \ce{K2SnX6}, simply by using the Goldschmidt tolerance factor, $t_G=(r_{\ce{K}}+r_{\ce{Sn}})/\sqrt{2}(r_{\ce{Sn}}+r_{\ce{X}})$, where $r_{\ce{K}}$, $r_{\ce{Sn}}$, and $r_{\ce{X}}$ are the Shannon ionic radii for K$^+$, Sn$^{4+}$, and X$^-$ ions, respectively.
Based on the fact that tolerance factor within the range of $0.8 < t_G < 1.0$ allows the formation of perovskite structure, we can expect in safe that all the three compounds \ce{K2SnX6} (X=I, Br, Cl) crystallize in the cubic perovskite phase due to their suitable tolerance factors of 0.88, 0.87, and 0.85 (Table~\ref{tab_lattice}).
On the other hand, Cai et al.~\cite{Cai17cm} used the octahedral factor, $t_o=r_{\ce{B}}/r_{\ce{X}}$, and radius ratio, $t_r=r_{\ce{A}}/(D_{\ce{XX}}-r_{\ce{X}})$, to empirically predict the formation and distortion of the crystalline structure in vacancy-ordered double perovskite \ce{A2BX6}, being $D_{\ce{XX}}$ the nearest neighbor X-X bond length calculated for cubic phase.
According to their survey for experimentally known \ce{A2BX6} compounds, a smaller $t_o$ disfavors a formation of \ce{BX6} octahedra, while a smaller $t_r$ favors a distortions of octahedra, lowering the symmetry of crystalline structure.
When the octahedral factor range within $0.29 < t_o <0.55$ and radius ratio within $0.87 < t_r <1.00$, \ce{A2BX6} stabilizes in cubic phase at room temperature.
As listed in Table~\ref{tab_lattice}, the octahedral factor and radius ratio decrease going from X = Cl to I, implying that as the ionic radius of the halogen anion increases the perovskite structure undergoes octahedral tilting and accordingly lowers its symmetry from the cubic structure to the lower-symmetry structure at room temperature for \ce{K2SnX6}.
Especially, with the calculated octahedral factor $t_o=0.39$ and radius ratio $t_r=0.89$ for \ce{K2SnCl6}, it can be concluded that it crystallizes in the stable cubic phase at room temperature as confirmed in the previous experiment~\cite{Boysen}.
It should be noted that although such considerations about structural factors of $t_G$, $t_o$, and $t_r$ could provide a qualitative prediction on the formation of perovskite structure and octahedral distortion, a quantitative description on phase stability should be given based on the lattice dynamics calculations.

Regarding the lattice constants, the PBE functional slightly overestimated compared with the experimental data~\cite{Boysen, Higashi}, with relative errors of less than 1\% for the cubic, tetragonal, and monoclinic phases of \ce{K2SnCl6} and \ce{K2SnBr6}.
The calculated lattice angle for the monoclinic phases were in good agreement with the experimental values~\cite{Boysen, Higashi}, with relative errors less than 0.3~\% (Table~\ref{tab_lattice}).
As increasing the ionic radius of the halogen anions, the lattice constants of all the phases increase and the lattice angle of the monoclinic phase deviates much from 90$^\circ$, indicating that the octahedral distortions become even more pronounced goning from X = Cl to Br and to I.
These can be attributed to the weakening of chemical bonds between the Sn and X atoms, subsequently increasing bond lengths and distorting octahedra.
Such changing tendencies in crystalline parameters coincide with the well-known fact that a perovskite with a smaller octahedral factor tends to form non-cubic structure with more distorted octahedra at room temperature.
Although there is a lack of experimental data for all the phases of \ce{K2SnI6}, we can expect that our work provides a reliable prediction for those.

\subsection{Electronic structures}
The electronic properties including the energy band structures and DOS can directly determine the performance of solar cells through the estimation of light absorption capability.
Thus, we calculated them, together with charge densities corresponding to the valence band maximum (VBM) and conduction band minimum (CBM) by using the PBE-GGA and hybrid HSE06 functionals with and without spin-orbit coupling (SOC) effect for all the phases of \ce{K2SnX6} (X = I, Br, Cl).
Figure~\ref{fig_band} shows the energy band structures calculated by using the HSE06 hybrid functional with and without SOC effect for each phase of \ce{K2SnX6} (X = I, Br, Cl).
For the cubic and tetragonal phases, all the perovskite compounds were found to have direct band gaps at $\Gamma$ point of the Brillouin zone (BZ), and such characteristics of direct band gap is consistent with previous DFT calculations in other types of vacancy-ordered double perovskites such as \ce{Cs2SnI6} and \ce{Rb2SnI6}~\cite{Cai17cm, LeeB}.
On the other hand, all the compounds in monoclinic phase were predicted to have indirect band gaps between CBM at $\Gamma$ point and VBM at $D$ point of BZ.
Interestingly, the energy band corresponding to VBM is almost dispersionless around $\Gamma$ point for the cubic and tetragonal phases and around $D$ point for the monoclinic phase, and from such flatness of valence band, we can expect that effective masses of holes are much larger than those of electrons.

\begin{table*}[!th]
\small
\caption{\label{tab_electric}Effective masses of electron ($m^*_e$) and hole ($m^*_h$), reduced mass ($m^*_r$), static and high-frequency dielectric constants ($\varepsilon_0$ and $\varepsilon_\infty$), exciton binding energies calculated by using the static ($E_b$) and high-frequency dielectric constants ($\tilde{E}_b$), and band gap ($E_g$) in \ce{K2SnX6} (X = Cl, Br, I), calculated with PBE and HSE06 functionals with and without SOC effect.}
\begin{tabular}{l@{\hspace{7pt}}l@{\hspace{7pt}}c@{\hspace{7pt}}c@{\hspace{7pt}}c@{\hspace{7pt}}c@{\hspace{7pt}}c@{\hspace{7pt}}r@{\hspace{7pt}}c@{\hspace{7pt}}r@{\hspace{7pt}}r@{\hspace{7pt}}c@{\hspace{7pt}}c@{\hspace{7pt}}c@{\hspace{7pt}}c@{\hspace{7pt}}c}
\hline
\multicolumn{1}{c}{} & \multicolumn{1}{c}{} & \multicolumn{3}{c}{PBE ($m_e$)} & & \multicolumn{2}{c}{PBE} & & \multicolumn{2}{c}{PBE (meV)} & & \multicolumn{4}{c}{$E_g$ (eV)}\\
\cline{3-5}  \cline{7-8} \cline{10-11} \cline{13-16}
material & phase & $m_e^*$ & $m_h^*$ & $m_r^*$ & & $\varepsilon_\infty$ & $\varepsilon_0$~~ & & $\tilde{E}_b$~~ & $E_b$~~ & & PBE & PBE+SOC & HSE06 & HSE06+SOC \\
\hline
\ce{K2SnCl6} & cub & 0.47 & 0.99 & 0.32 & & 2.69 & 6.28 & & 599.9 & 110.1 & & 2.41 & 2.40 & 3.43 & 3.36 \\
             & tet & 0.50 & 1.06 & 0.34 & & 2.71 & 6.45 & & 632.0 & 111.6 & & 2.45 & 2.44 & 3.51 & 3.49 \\
             & mono & 1.01 & 1.69 & 0.63 & & 2.75 & 8.33 & & 1300.8 & 123.8 & & 2.96 & 2.93 & 4.05 & 4.04 \\
\hline
\ce{K2SnBr6} & cub & 0.33 & 0.83 & 0.24 & & 3.29 & 7.36 & & 297.5 & 59.4 & & 1.01 & 0.92 & 1.81 & 1.65 \\
             & tet & 0.46 & 0.83 & 0.29 & & 3.48 & 11.45 & & 330.6 & 30.6 & & 1.56 & 1.40 & 2.50 & 2.32 \\
             & mono & 0.72 & 1.15 & 0.44 & & 3.86 & 12.52 & & 403.5 & 38.3 & & 1.83 & 1.77 & 2.68 & 2.57 \\
\hline
\ce{K2SnI6} & cub & 0.17 & 0.46 & 0.12 & & 4.95 & 13.76 & & 68.4 & 8.9 & & 0.05 & 0.04 & 0.52 & 0.31 \\
            & tet & 0.39 & 0.69 & 0.25 & & 5.54 & 15.41 & & 108.4 & 14.0 & & 0.44 & 0.32 & 0.96 & 0.74 \\
            & mono & 0.58 & 0.78 & 0.33 & & 9.59 & 17.20 & & 104.0 & 15.3 & & 0.81 & 0.69 & 1.40 & 1.16 \\
\hline             
\end{tabular}
\end{table*}
\begin{figure}[!th]
\begin{center}
\includegraphics[clip=true,scale=0.45]{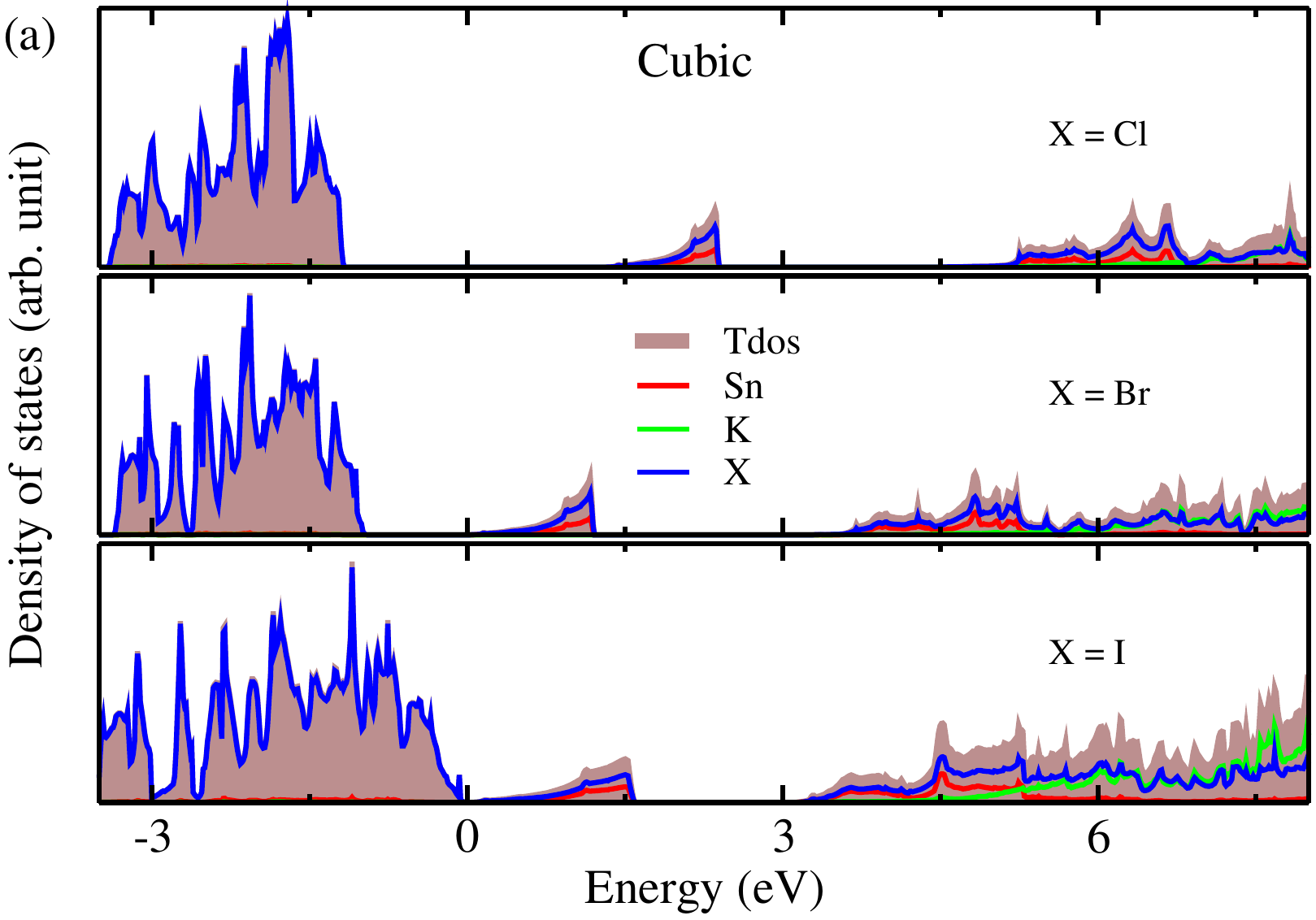} \\
\includegraphics[clip=true,scale=0.45]{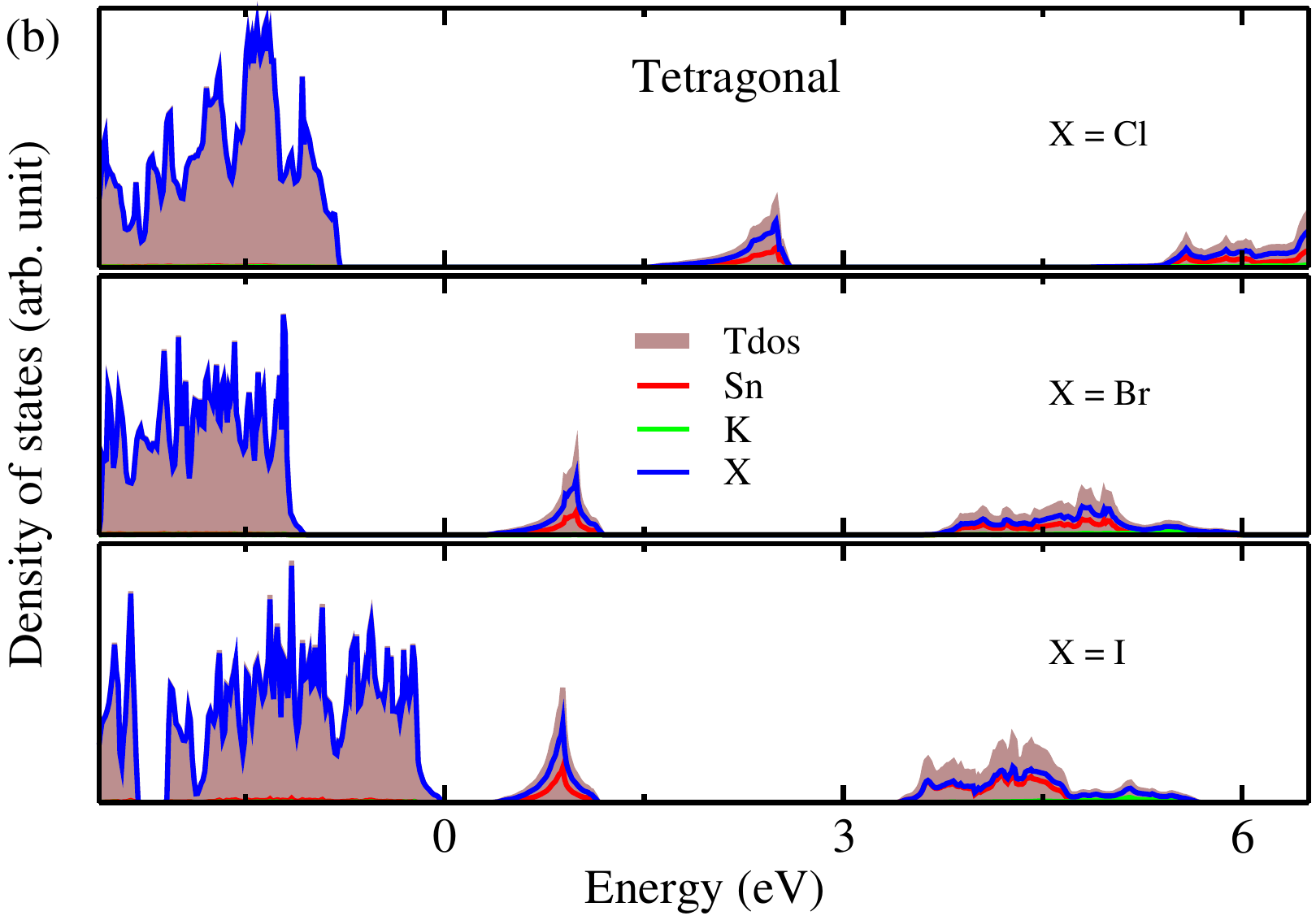} \\
\includegraphics[clip=true,scale=0.45]{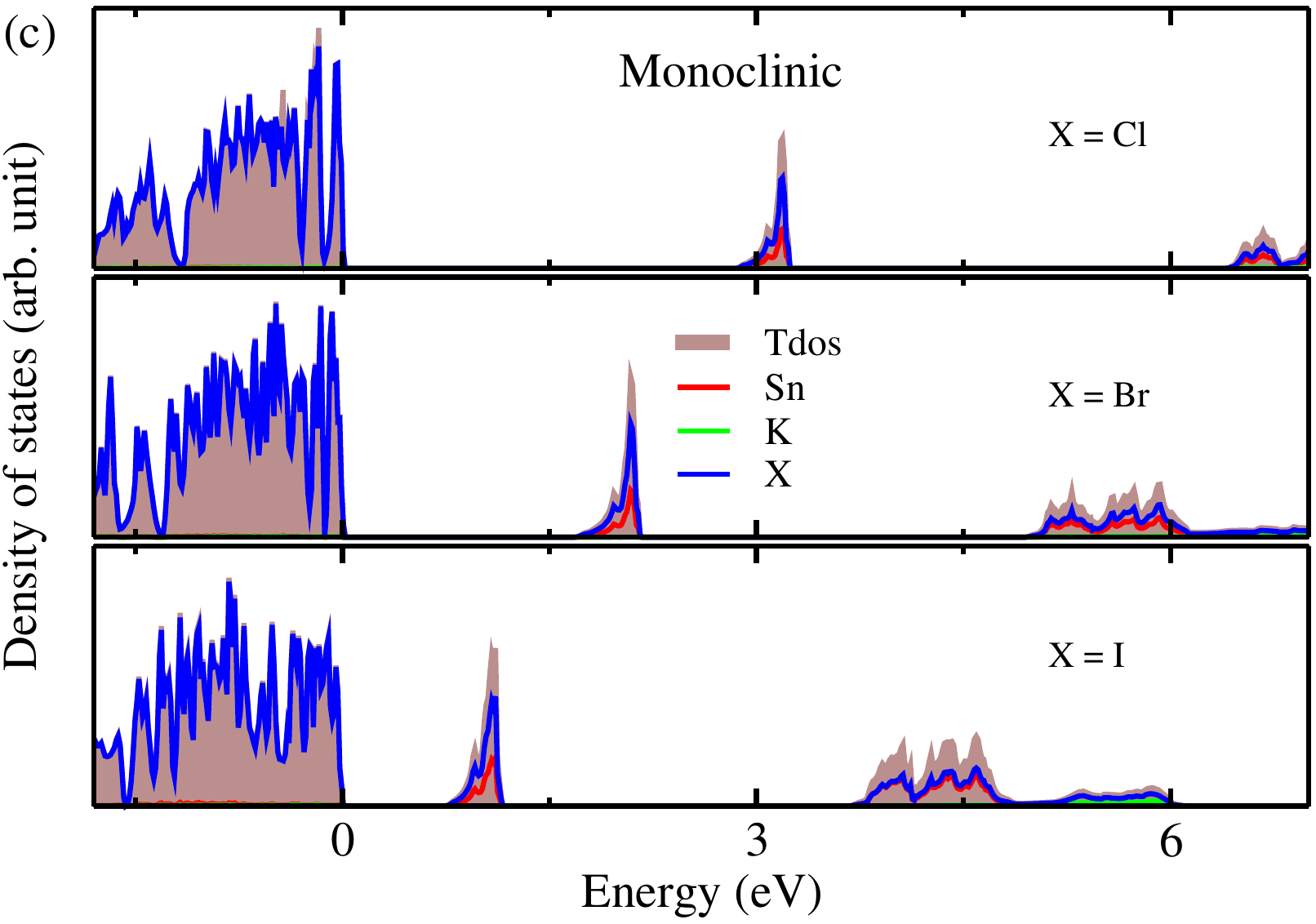}
\end{center}
\caption{\label{fig_dos}Total and atomic resolved electronic density of states in \ce{K2SnX6} (X = I, Br, Cl) in (a) cubic, (b) tetragonal and (c) monoclinic phases, calculated by HSE functional with SOC effect.}
\end{figure}

In Table~\ref{tab_electric}, we list the band gaps calculated by using PBE and HSE06 functionals with and without SOC effect for the cubic, tetragonal, and monoclinic phases of \ce{K2SnX6} (X = I, Br, Cl).
As like for other insulating compounds, the HSE06 calculations were found to widen the band gaps compared with PBE calculations.
As shown in Figure~\ref{fig_band}, when considering the SOC effect, the valence (conduction) bands were found to be slightly pushed up (down) in comparison with those without the SOC effect, resulting in narrowing the band gaps for all the phases of \ce{K2SnX6}.
It should be noted that the SOC effect becomes weaker going from heavier halogen element of X = I to Cl, as the band gap difference between with and without SOC effect becomes smaller for all the phases.
By considering the fact that HSE06 + SOC calculation can provide reasonable band gap in good accordance with experiment, it can be said that \ce{K2SnI6} in the monoclinic phase and \ce{K2SnBr6} in the cubic phase are suitable for application of light-absorber due to their proper band gaps of 1.16 and 1.65 eV calculated by HSE06 + SOC method.
On the other hand, the band gaps of \ce{K2SnCl6} in cubic, tetragonal, and monoclinic phases were estimated by HSE06 + SOC method to be 3.36, 3.49 and 4.04 eV respectively, implying that the chlorine-based double perovskites are not applicable for light-absorbers but might be appropriate for charge carrier conducting materials.
Meanwhile, for the cubic and tetragonal phases of \ce{K2SnI6} compound, the HSE06 + SOC calculations yielded the smaller band gaps of 0.31 and 0.74 eV, indicating that those can be useful for applications of infrared emitting diodes.

\begin{figure*}[!th]
\begin{center}
\begin{tabular}{cccc}
& cubic & tetragonal & monoclinic \\
VBM & \includegraphics[clip=true,scale=0.13]{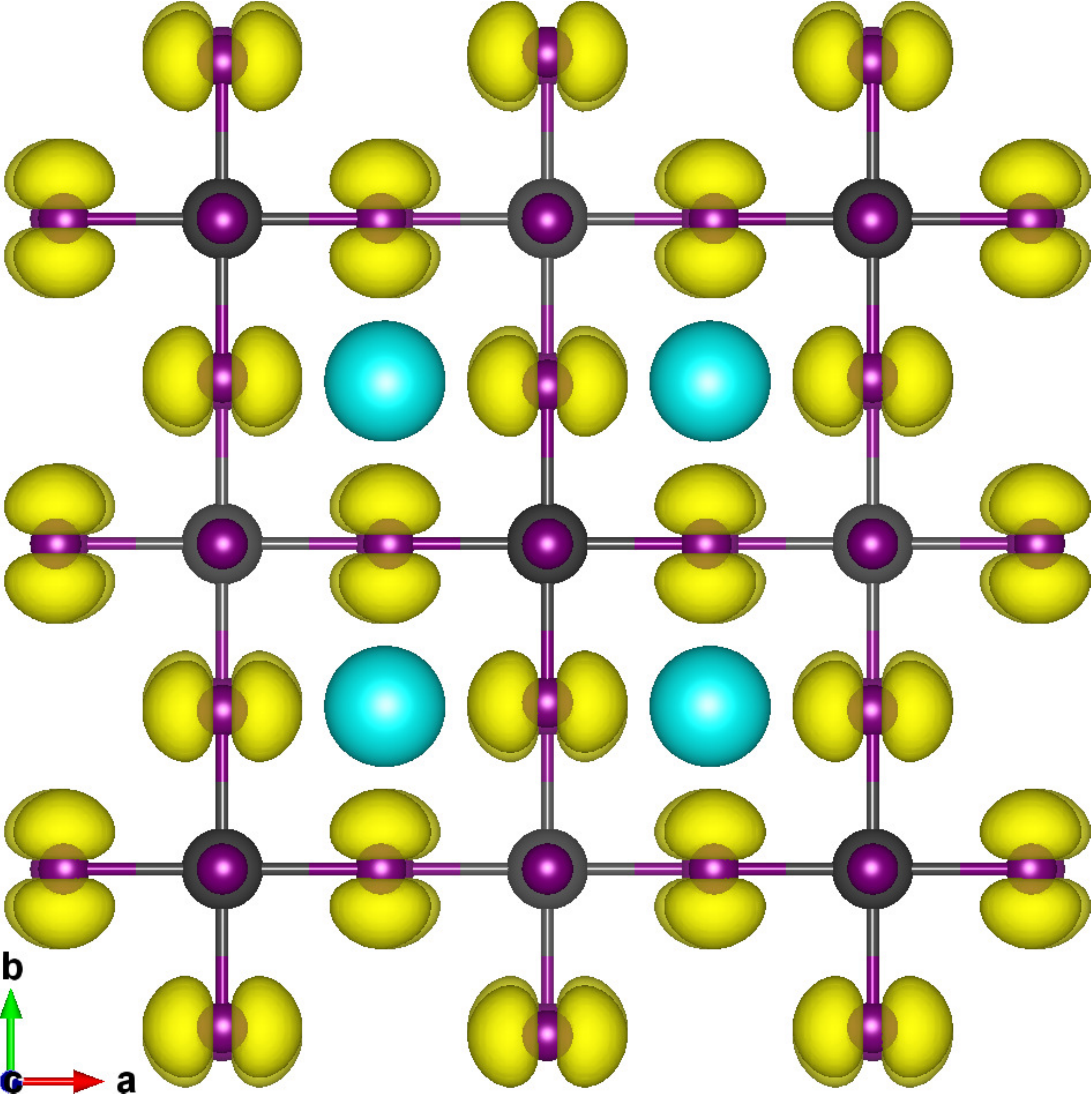} &
\includegraphics[clip=true,scale=0.14]{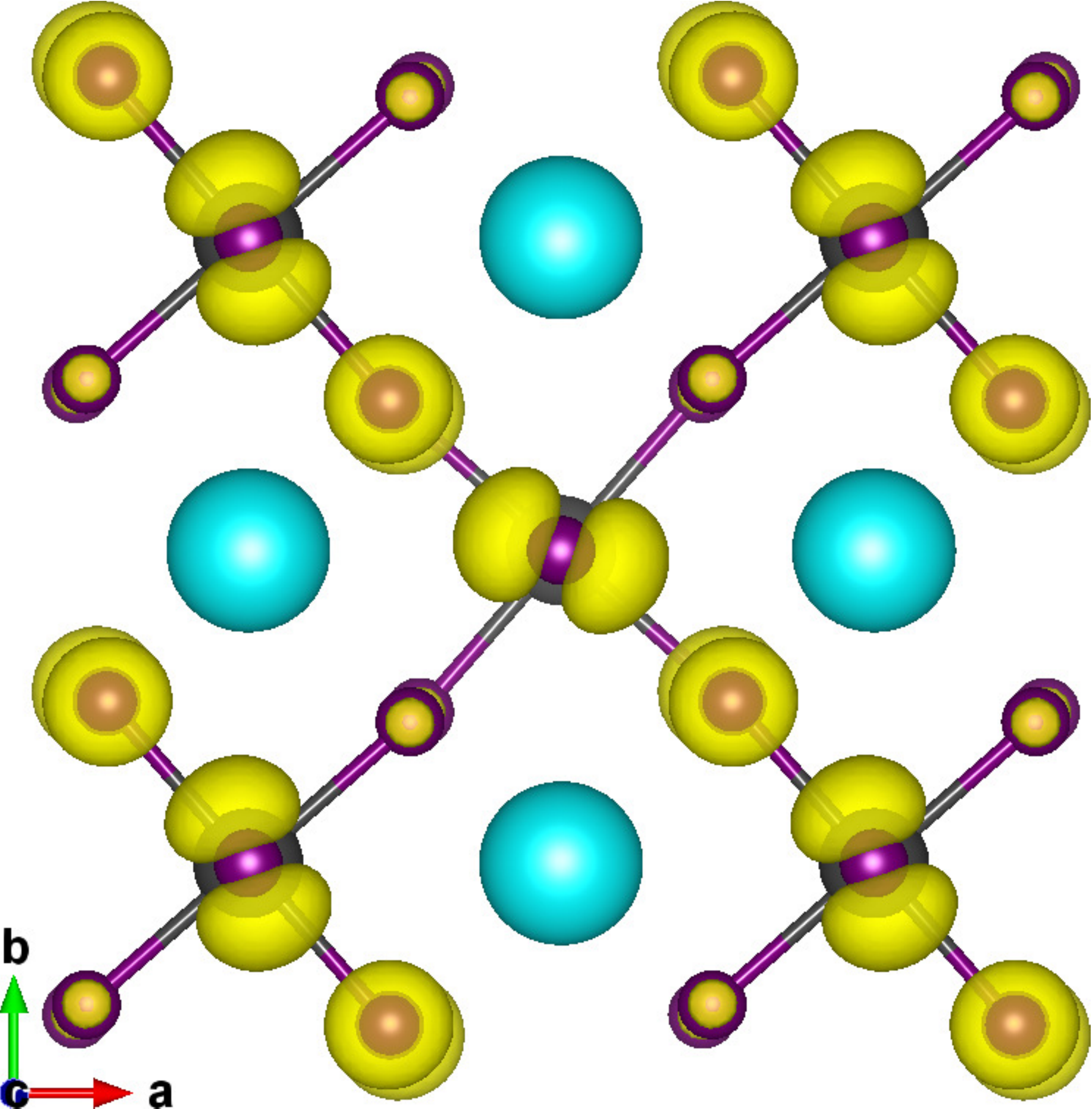} &
\includegraphics[clip=true,scale=0.14]{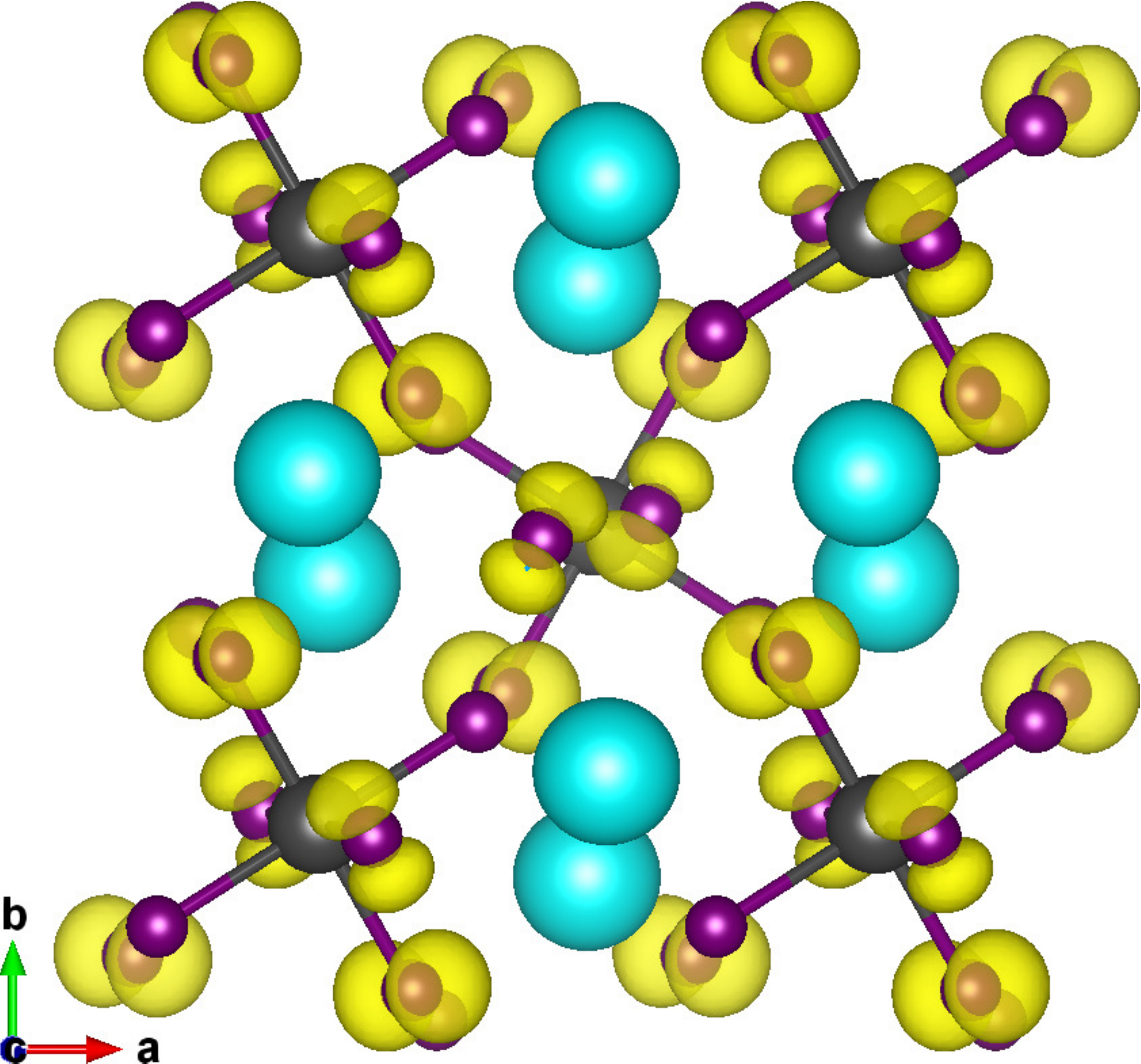} \\
& & & \\
CBM & \includegraphics[clip=true,scale=0.13]{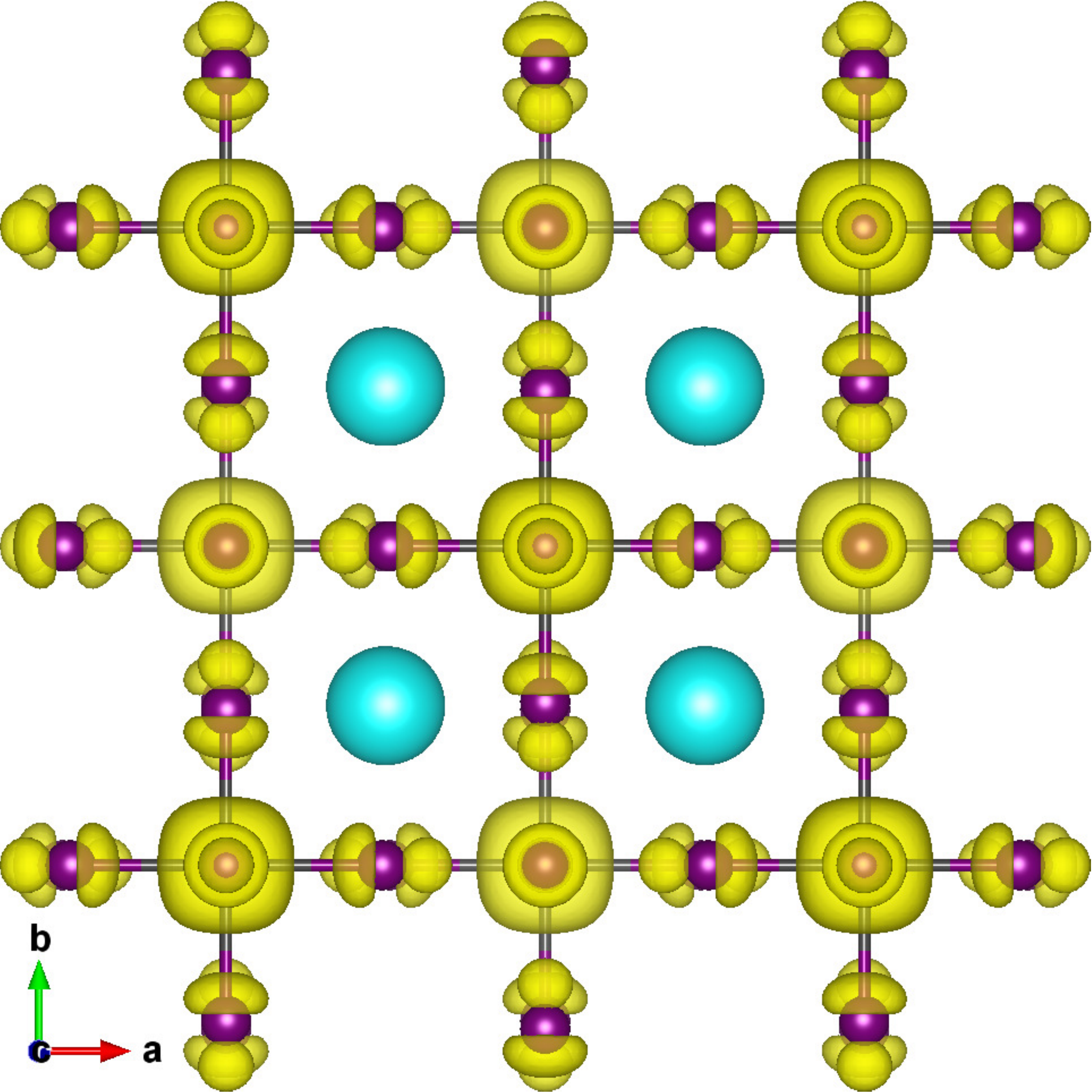} &
\includegraphics[clip=true,scale=0.14]{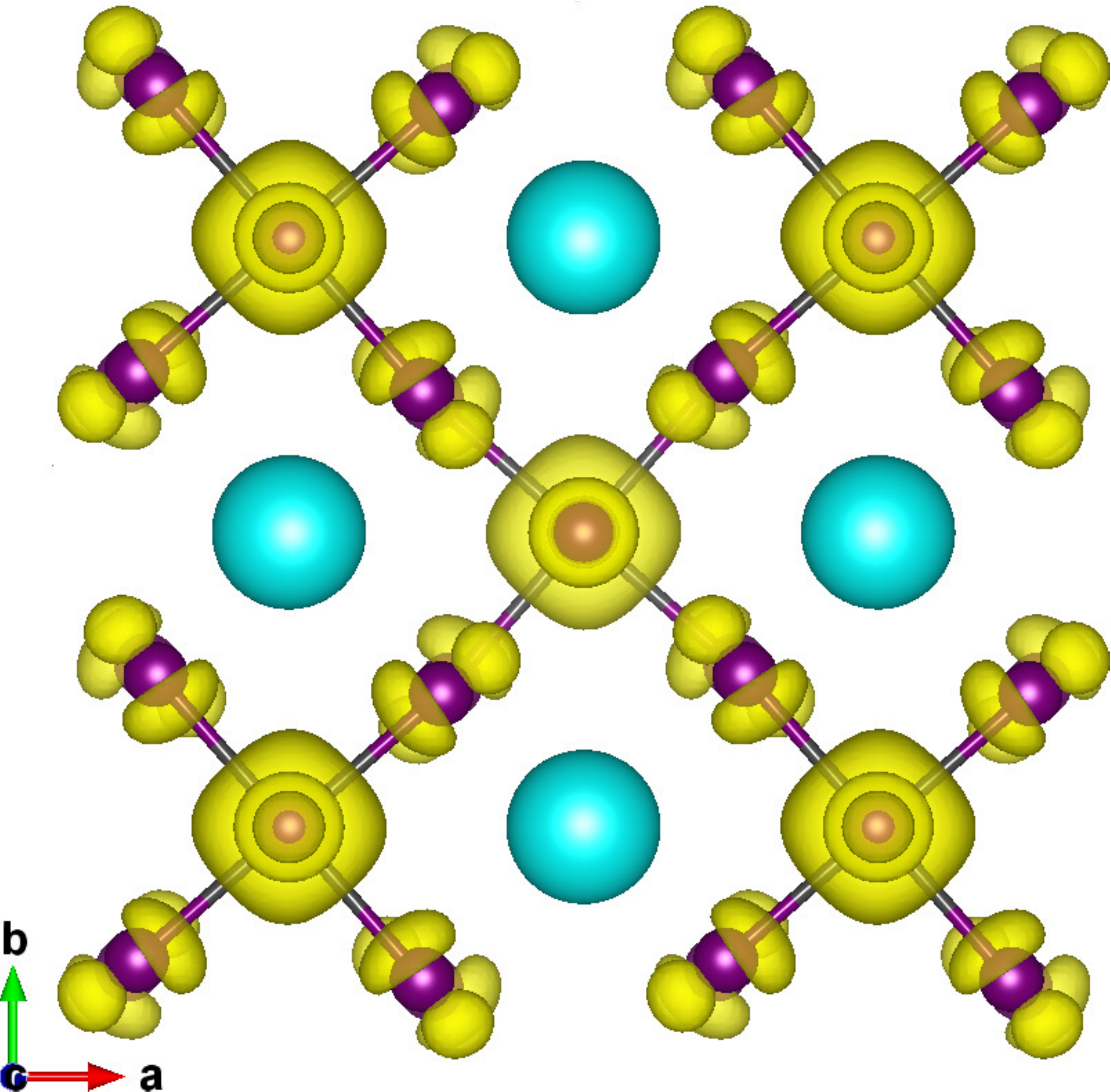} &
\includegraphics[clip=true,scale=0.14]{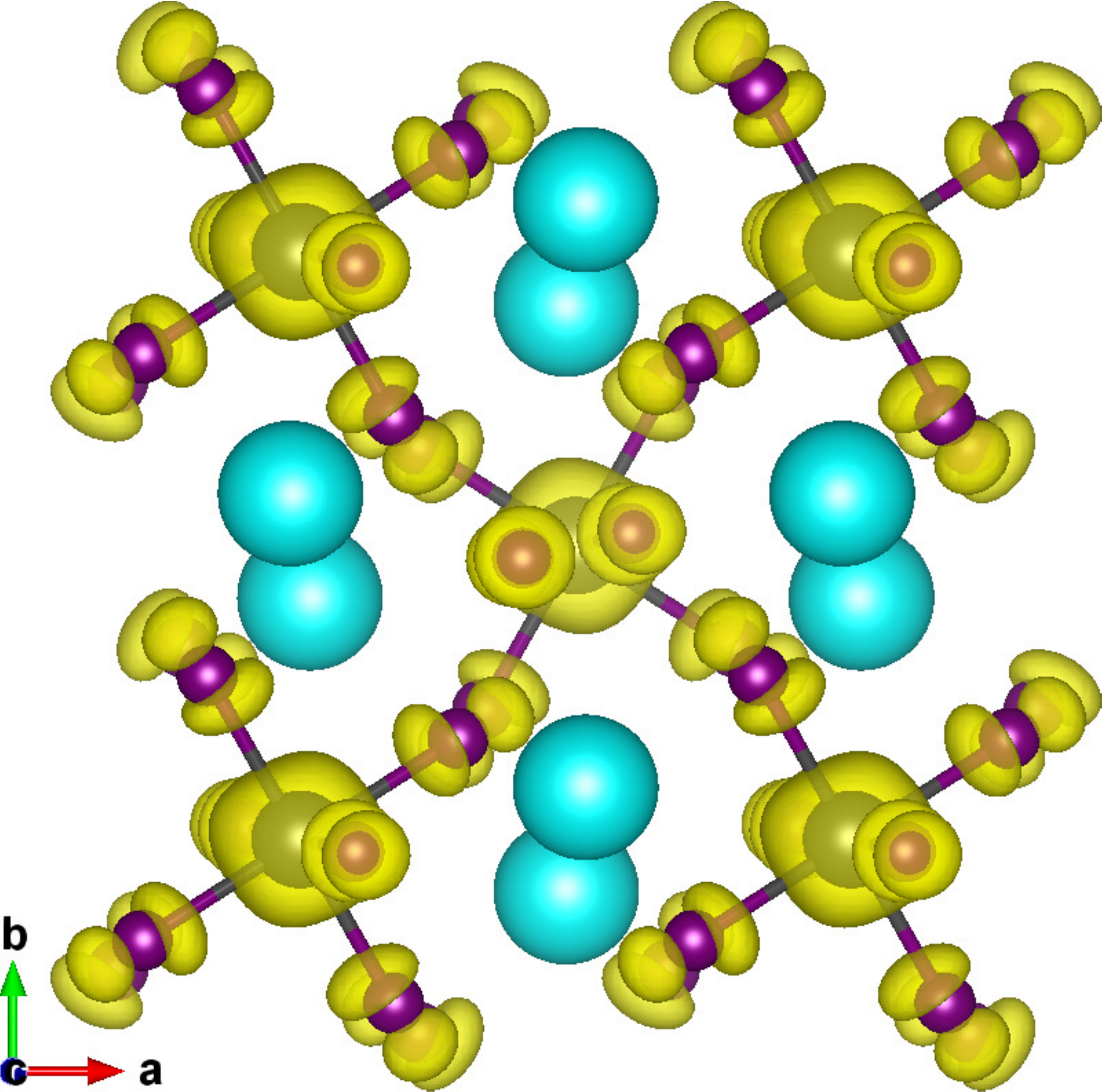} \\
\end{tabular} 
\end{center}
\caption{\label{fig_charge}Isosurface plot of charge density corresponding to the valence band maximum (VBM) and conduction band minimum (CBM) at the value of 0.02~$|e|$/\AA$^3$ in \ce{K2SnI6} in cubic, tetragonal, and monoclinic phases, calculated by PBE functional. Grey, purple and green balls represent Sn, I and K atoms.}
\end{figure*}
The calculated band gaps display a distinct variation tendency with respect to the choice of halogen atom, such that the band gaps decrease systematically as the ionic radius of halogen anion increases for all the phases.
Such variation tendency agrees well with the previous calculations for the hybrid organic-inorganic, all-inorganic, and vacancy-ordered double perovskites~\cite{Cai17cm, LeeB, jong19ic, Jong16prb, jong2017jps}, which can be understood through the analysis of total and atomic resolved DOS (see Figure~\ref{fig_dos}).
As can be seen in Figure~\ref{fig_charge}, VBM is derived from the $p$ orbitals of halide anion, while CBM is characterized by antibonding between the Sn $s$ and the halide $p$ orbitals.
Therefore, as increasing the size of ionic radius of halide anion and thus decreasing its electronegativity as going from Cl to I, VBM becomes higher while CBM lower, resulting in the decrease of band gap~\cite{Cai17cm, Jong16prb}.
It should be noted that lowering the symmetry from cubic to monoclinic increases the band gap for all the compounds \ce{K2SnX6} (X = I, Br, Cl) (see Table~\ref{tab_electric}), which correlates with the fact that as the symmetry lowers, the degree of octahedral distortion increases, resulting in the decrease of bonding strength between the neighboring halide anions, thus narrowing of valence bands and the increase of band gap.

\begin{figure*}[!t]
\small
\begin{center}
\begin{tabular}{ccc}
\includegraphics[clip=true,scale=0.26]{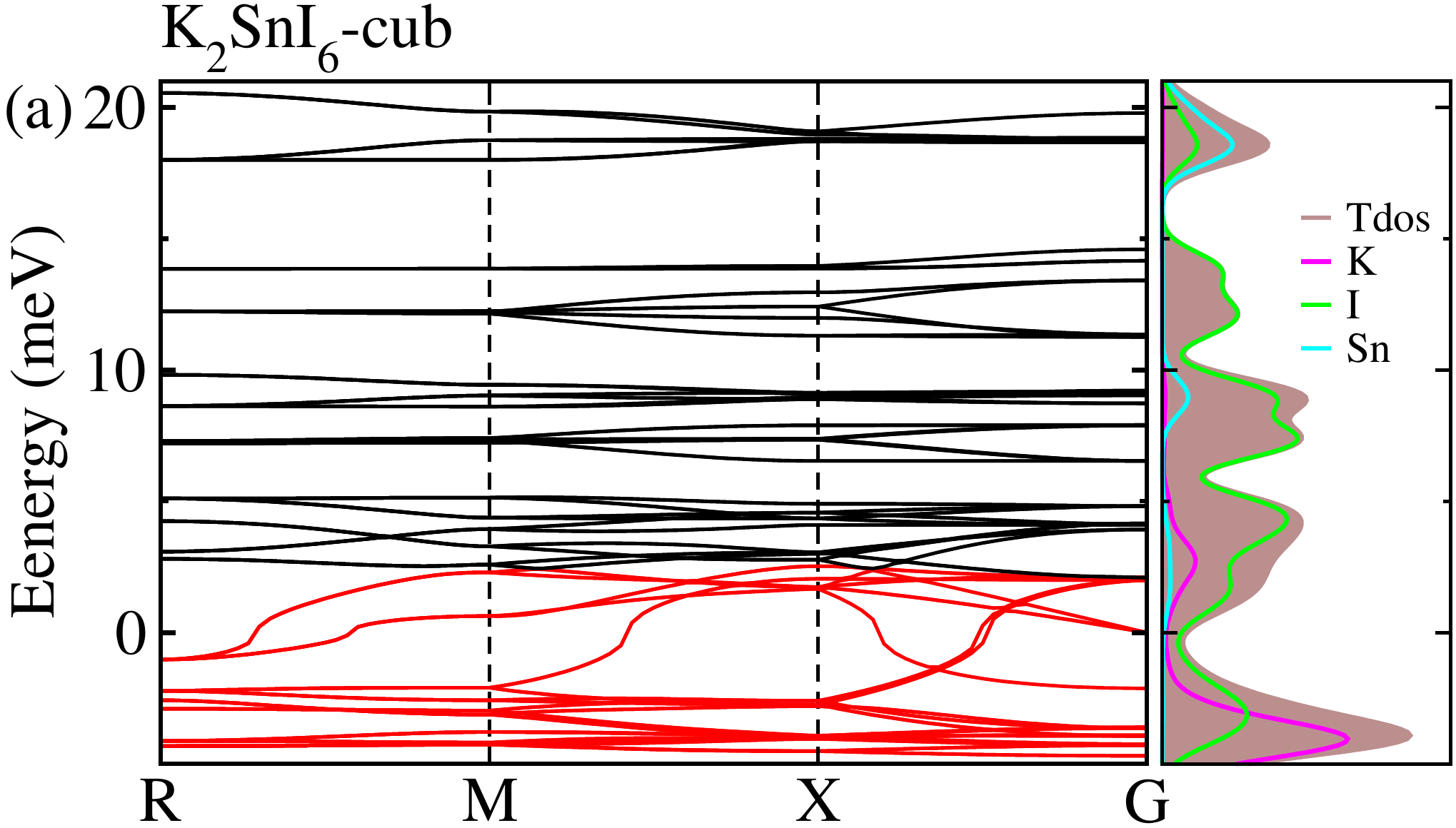} &
\includegraphics[clip=true,scale=0.26]{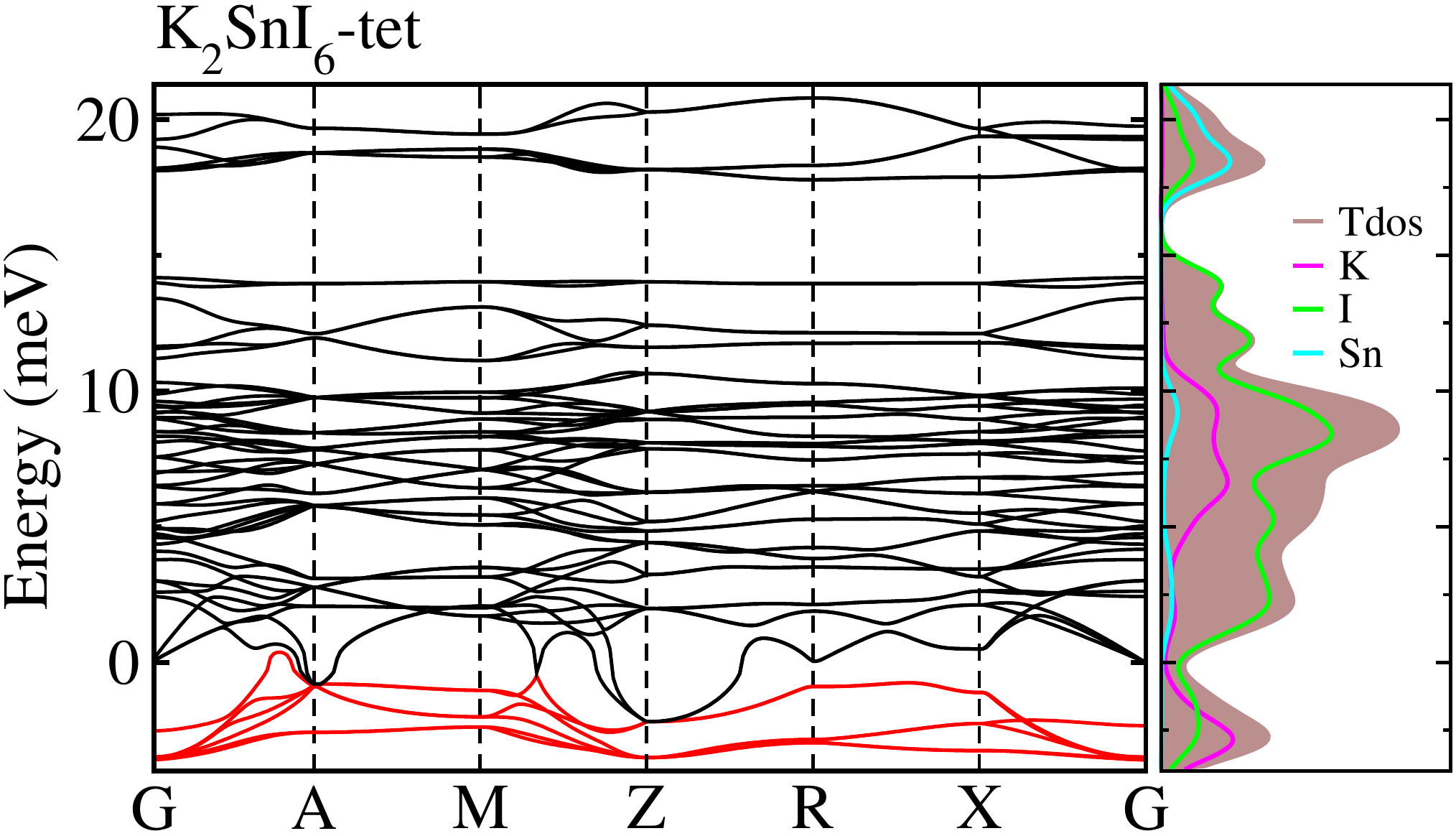} &
\includegraphics[clip=true,scale=0.26]{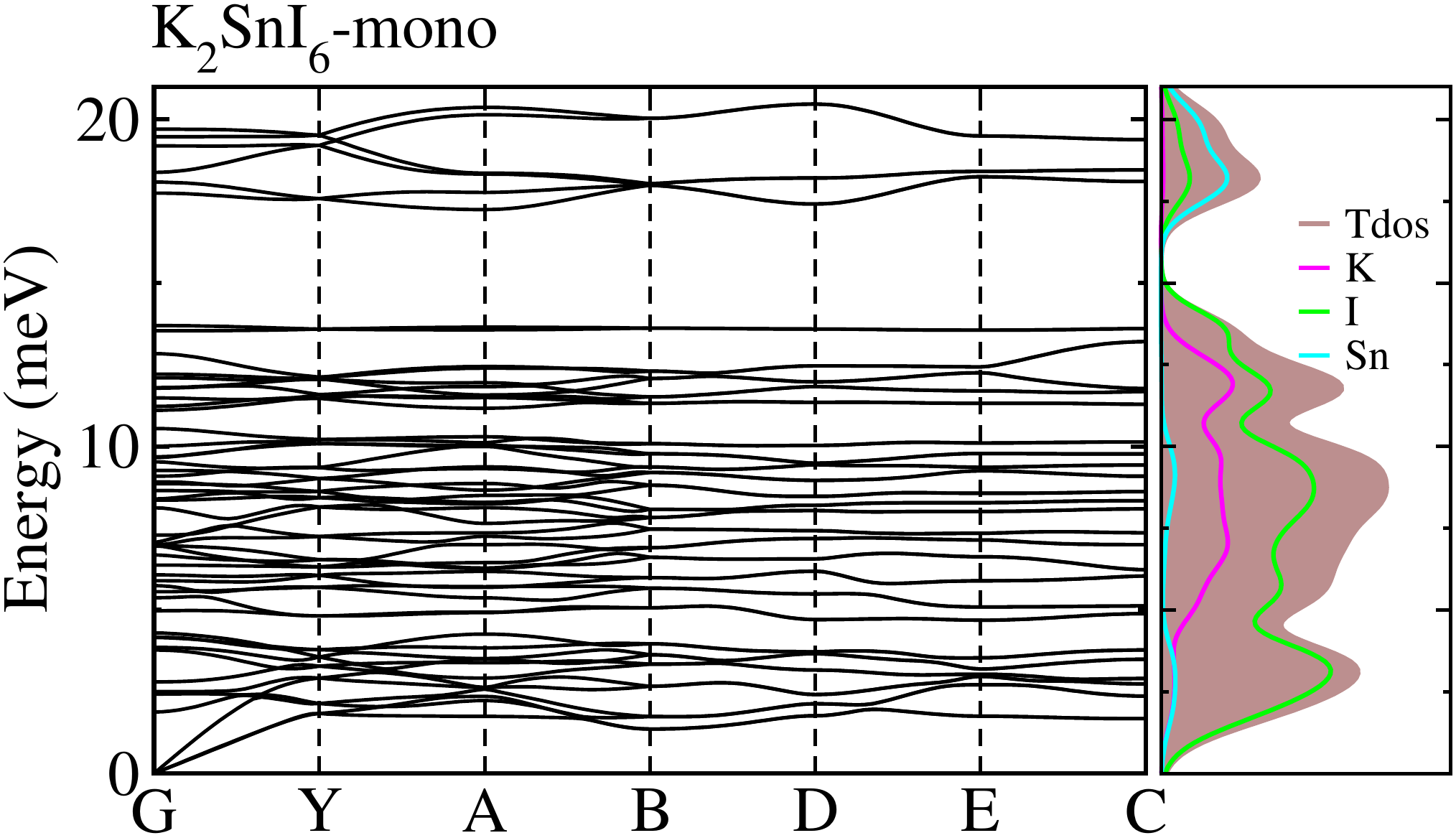} \\
\includegraphics[clip=true,scale=0.26]{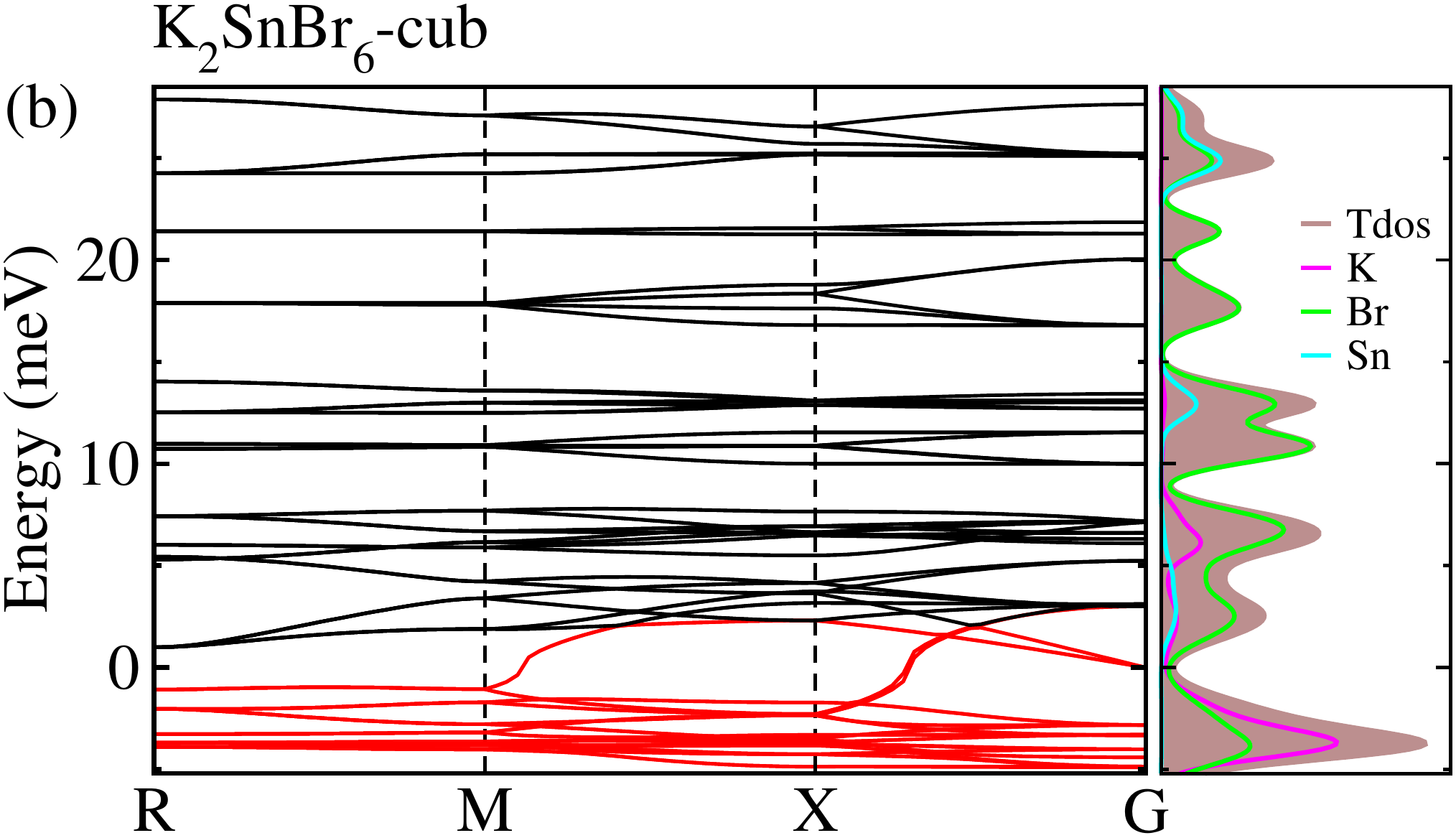} &
\includegraphics[clip=true,scale=0.26]{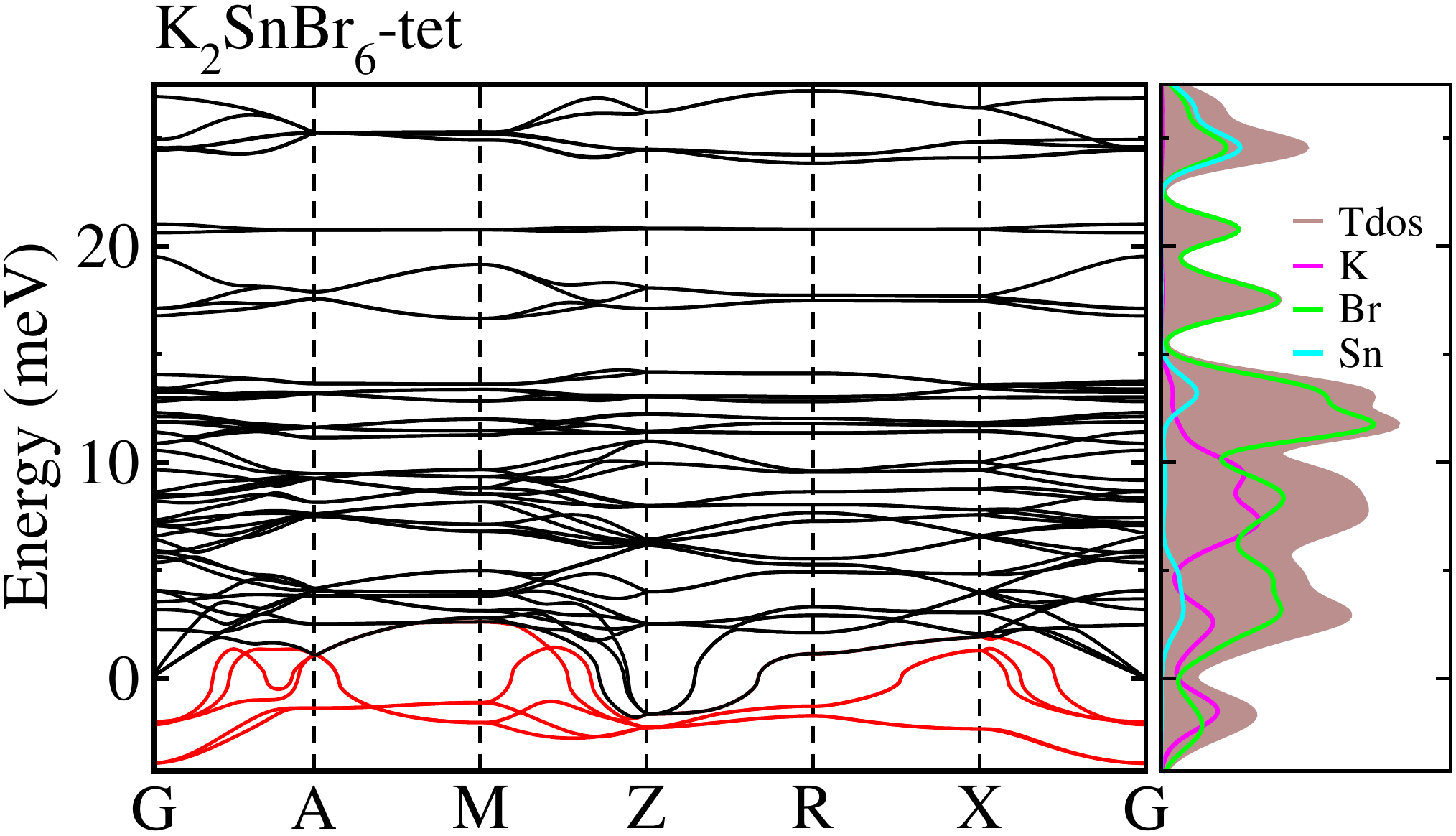} &
\includegraphics[clip=true,scale=0.26]{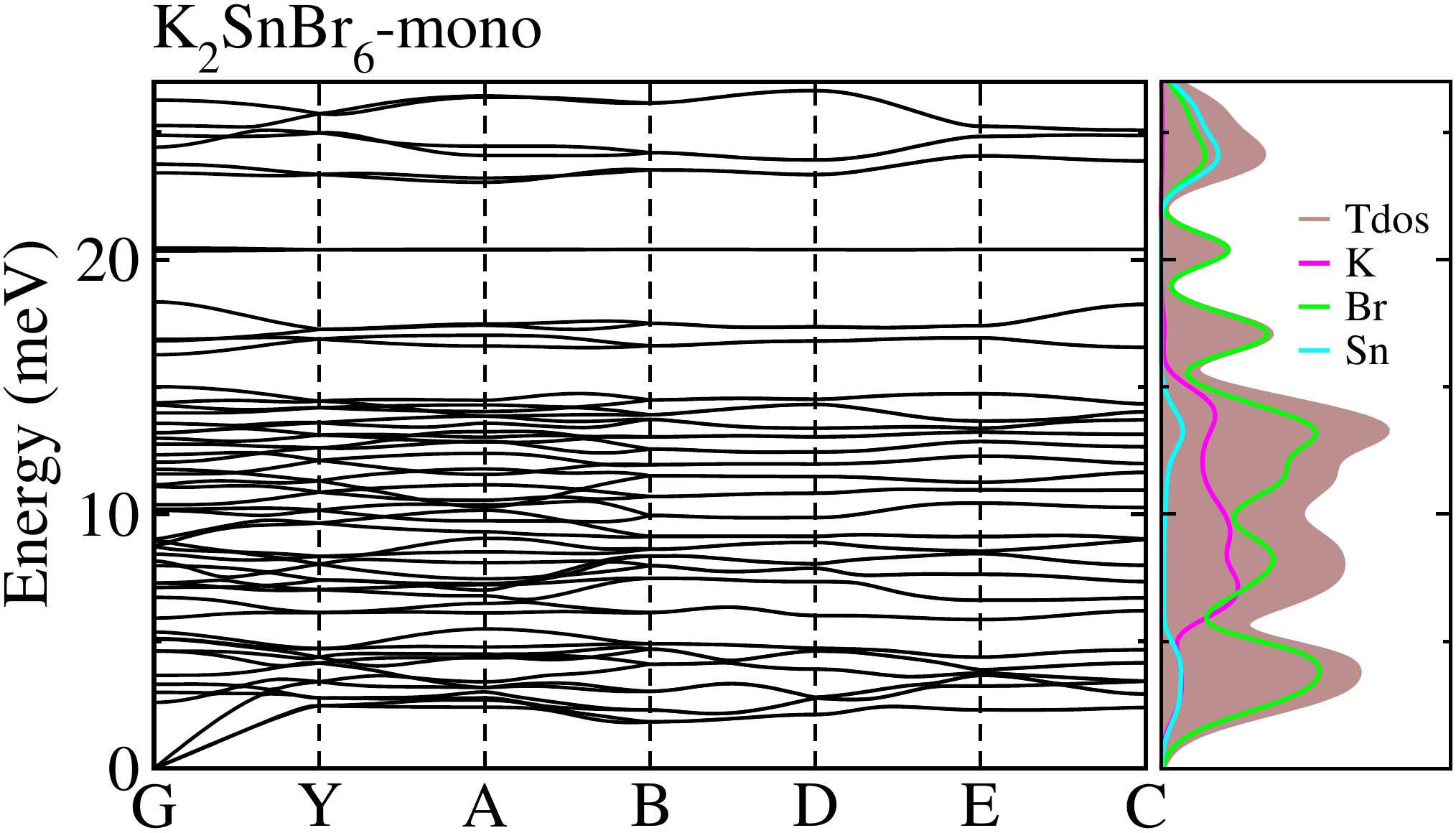} \\
\includegraphics[clip=true,scale=0.26]{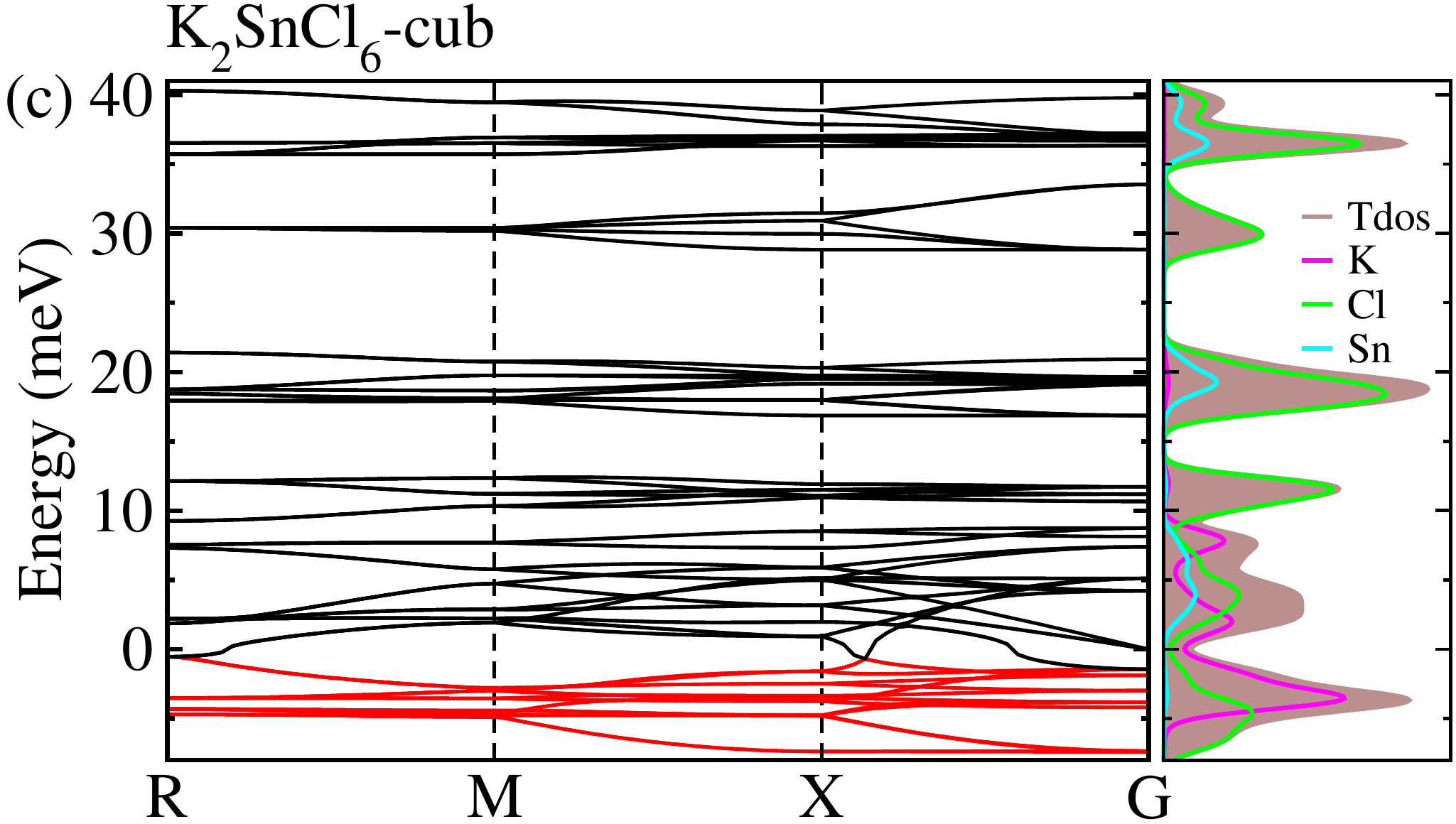} &
\includegraphics[clip=true,scale=0.26]{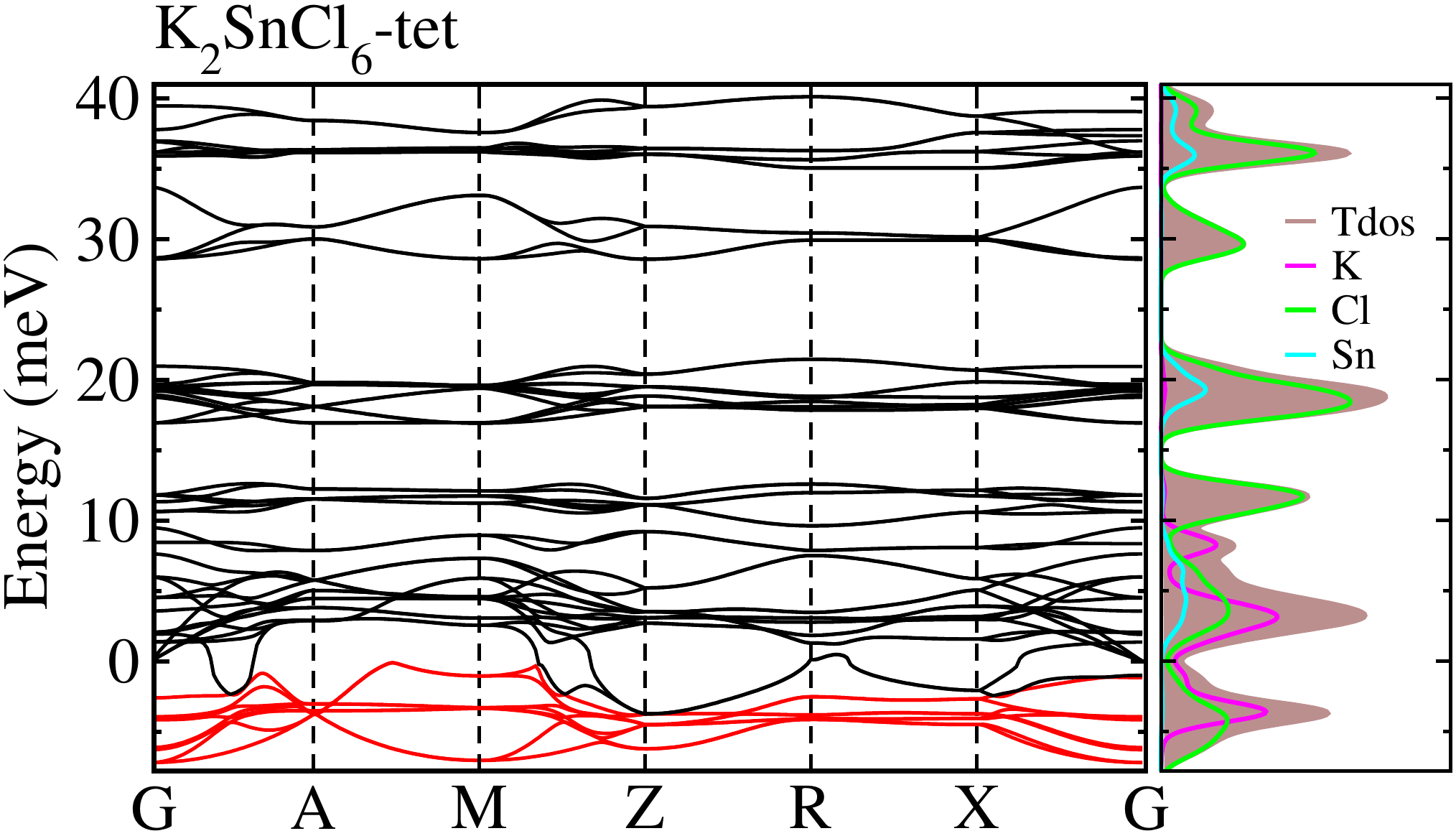} &
\includegraphics[clip=true,scale=0.26]{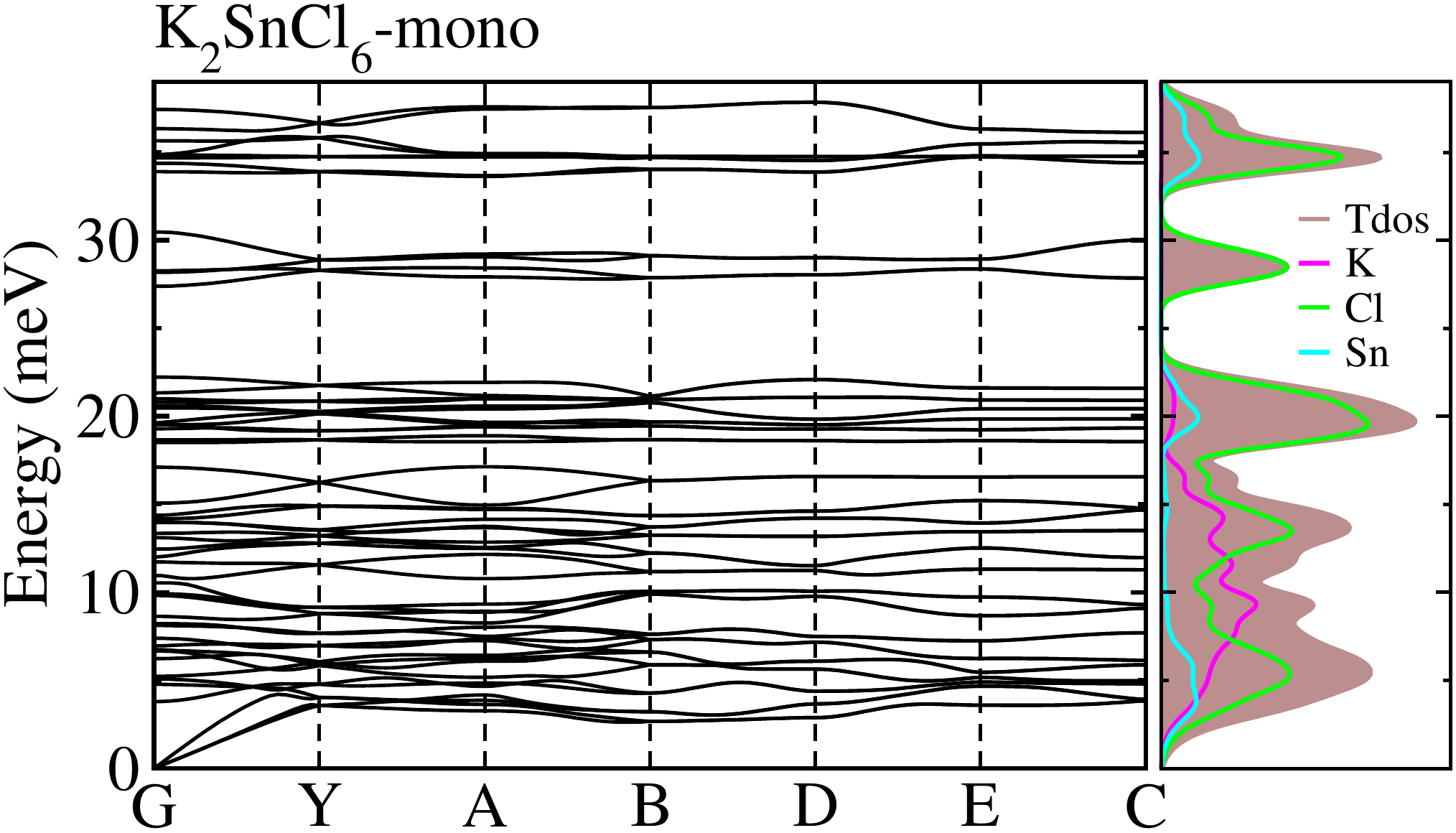} \\
\end{tabular} 
\end{center}
\caption{\label{fig_phonon}Phonon dispersion curves and atomic resolved density of states in (a) \ce{K2SnI6}, (b) \ce{K2SnBr6} and (c) \ce{K2SnCl6} in cubic (left), tetragonal (middle) and monoclinic (right) phases. Red color lines display the anharmonic phonon modes with negative phonon energies.}
\end{figure*}

\subsection{Optical properties}
In a next step, we further investigated various optical properties of \ce{K2SnX6} (X = I, Br, Cl) compounds, including effective masses of electrons and holes, static and high-frequency dielectric constants, exciton binding energies, and light-absorption coefficients by use of the PBE functional, so as to assess their potentiality of using as light-absorbers (Table~\ref{tab_electric}).
The effective masses of electron and hole were observed to have variation tendency similar to the case of band gap.
That is, for a given compound, the effective masses of electrons and holes increase as moving from the cubic to the monoclinic phases, whereas for a given phase they decrease going from Cl to I.
This straightly indicates that changing from lower- to higher-symmetry phase as well as going from X = Cl to I, \ce{K2SnX6} is expected to possess high mobility of charge carriers owing to the smaller effective masses of electron and hole.
In addition, as confirmed in other types of double perovskite\cite{Cai17cm, LeeB}, the effective masses of holes were found to be larger than those of electrons for all the phases of \ce{K2SnX6} (X = I, Br, Cl).
Our calculated effective masses of $m_e^*=0.17m_e$ and $m_h^*=0.46m_e$ for the cubic \ce{K2SnI6} are smaller than the previous calculations of $m_e^*=0.29m_e$ and $m_h^*=1.34m_e$ for the cubic \ce{Rb2SnI6}, and $m_e^*=0.33m_e$ and $m_h^*=1.50m_e$ for the cubic \ce{Cs2SnI6}~\cite{Cai17cm}, which agrees well with Cai's report~\cite{Cai17cm} where they concluded that reducing the size of the A-site cation leads to a decrease in effective masses of both electron and hole for \ce{A2BX6}.

We next consider the dielectric constant which plays an important role in the assessment of optical properties.
In this work, we calculated two kinds of dielectric constants such as high-frequency ($\varepsilon_\infty$) and static ($\varepsilon_0$) ones, of which the former was extracted from frequency-dependent macroscopic dielectric functions calculated using the DFPT approach and the latter was estimated by post-processing the phonon dispersion properties.
The calculated dielectric constants were shown to have a variation tendency slightly different from the cases of band gap and effective masses.
For a given perovskite compound, the dielectric constant increases going from the cubic to the monoclinic phases in accordance with the cases of band gap and effective masses, whereas for a given symmetry, they decrease as reducing the ionic radius of halogen anion on the contrary to the former cases.
It was eventually found that both lowering the symmetry and increasing the ionic radius of halogen anion increase the dielectric constants for \ce{K2SnX6} (X = I, Br, Cl).

By use of the calculated effective masses of charge carriers and dielectric constants, we obtained exciton binding energies that play a key role in discriminating whether electrons and holes behave as bound excitons or free charge carriers.
In Table~\ref{tab_electric}, we listed two types of exciton binding energies of $\tilde{E}_b$ and $E_b$ which are respectively calculated with the high-frequency ($\varepsilon_\infty$) and static ($\varepsilon_0$) dielectric constants.
It is clear that $E_b$ is reduced by a factor of at least 6 compared to $\tilde{E}_b$, because by using the static dielectric constant, phonon processes contribute to screening the electrostatic interactions between electrons and holes, and subsequently, weakening their binding energies.
For the cubic, tetragonal, and monoclinic phases of \ce{K2SnI6}, the exciton binding energies $E_b$ were estimated to be 8.9, 12.0, and 15.3 meV, which are obviously smaller than the value of 45-50 meV for the cubic \ce{MAPbI3}~\cite{Jong16prb, jong2017jps}, while $E_b$ of \ce{K2SnCl6} was determined to be about 2 times larger than the value of cubic \ce{MAPbI3}.
At the end, we emphasize that the calculated exciton binding energies have the same variation tendency to the band gap and effective masses, according to the change of phase symmetry and the size of halogen anion.
Light-absorption coefficients were obtained by using the PBE calculated frequency-dependent dielectric constants, showing that the absorption onset was gradually shifted to the blue color as the size of halide anion decreases for a given phase, and as the symmetry of phase lowers for a given chemistry (see Figure S2).

\subsection{Phase stability}
As a final step, we investigated phonon dispersion with phonon total and atomic resolved DOS for the cubic, tetragonal, and monoclinic phases for \ce{K2SnX6} (X = I, Br, Cl) by using the DFPT method with the PBE functional without SOC effect.
As shown in Figure~\ref{fig_phonon}, the phonon dispersion curves of the cubic and tetragonal phases presented the anharmonic phonon modes with the negative phonon energies, indicating their dynamic instabilities at low temperature, whereas the monoclinic phases did not exhibit the anharmonic feature, meaning that \ce{K2SnX6} (X = I, Br, Cl) can stabilize in monoclinic phase at low temperature.
It is clear from the atomic resolved phonon DOS that as K and X atoms contribute to the phonon DOS corresponding to the anharmonic phonon modes, the dynamic instabilities in the cubic and tetragonal phases are attributed to the vibrations of K and X atoms.

\begin{figure}[!th]
\includegraphics[clip=true,scale=0.45]{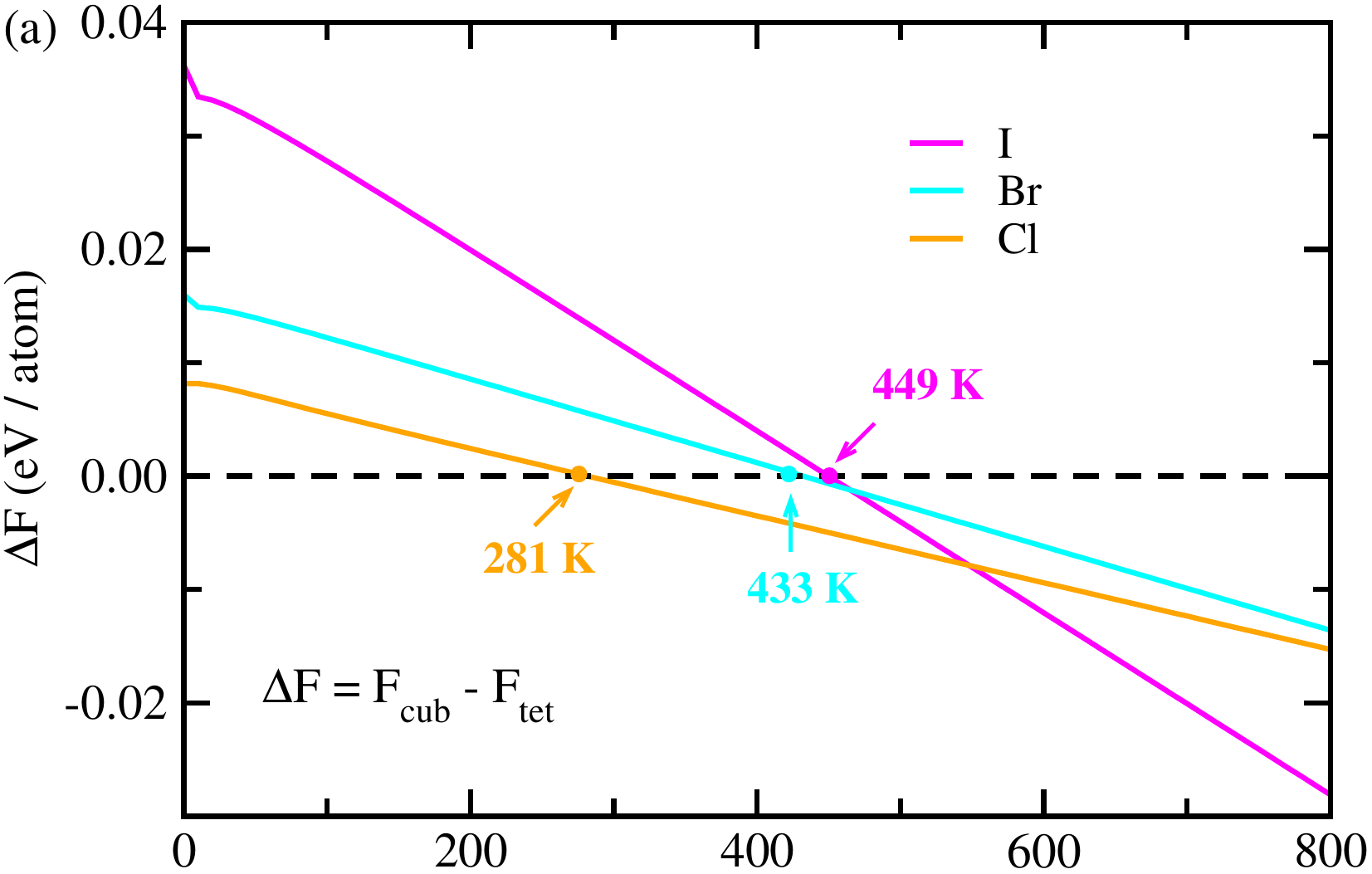} \\
\includegraphics[clip=true,scale=0.45]{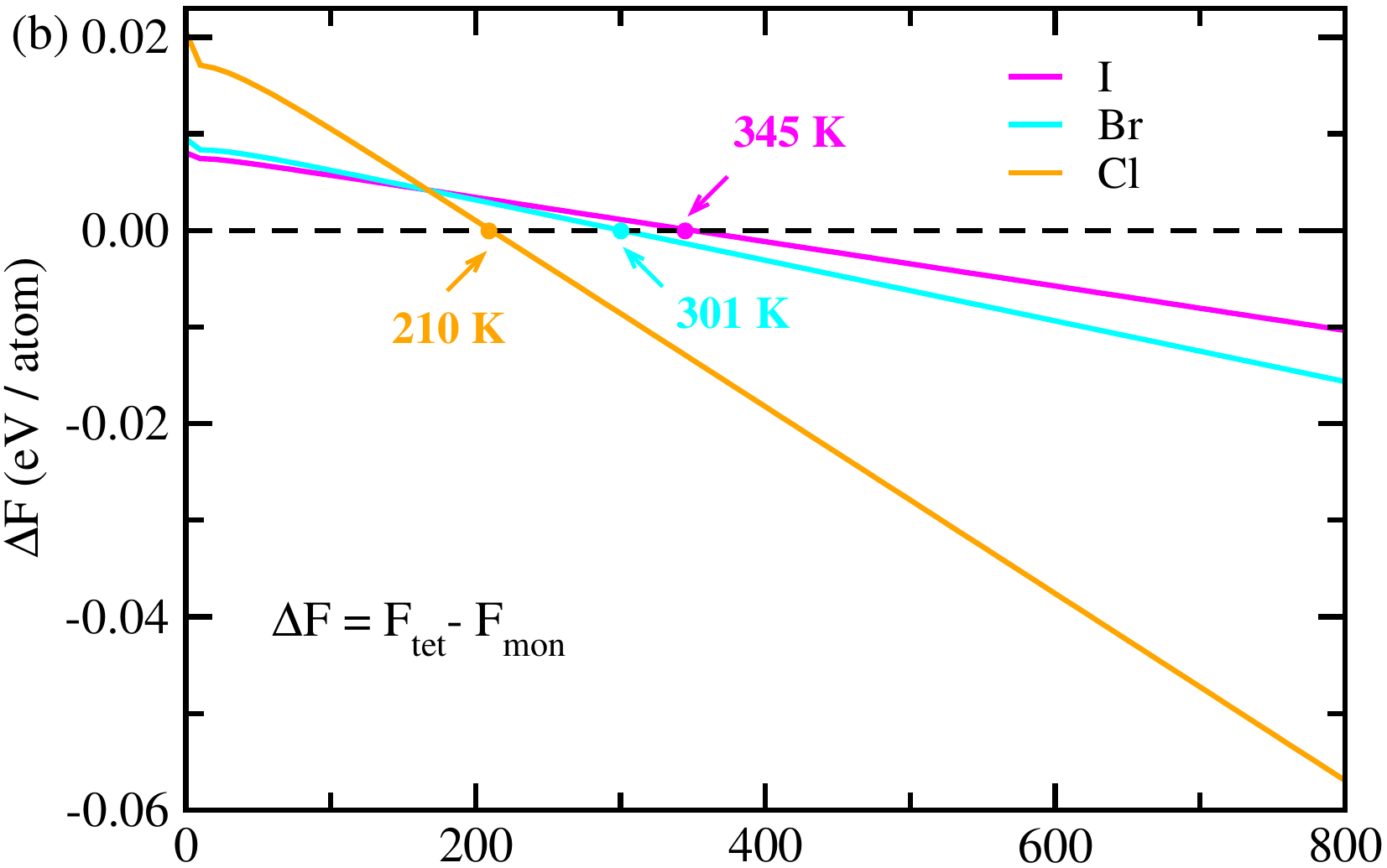}
\caption{\label{fig_fenergy}Helmholtz free energy differences of (a) cubic from tetragonal and (b) tetragonal from monoclinic phases in \ce{K2SnX6} (X = I, Br, Cl).}
\end{figure}
By post-processing the phonon DOS, we finally calculated the constant-volume Helmholtz free energies of the cubic, tetragonal, and monoclinic phases for \ce{K2SnX6} (X = I, Br, Cl) as increasing temperature from 0 to 1000 K with an interval of 10 K.
The phase transition temperatures were estimated by the free energy differences of cubic from tetragonal and of tetragonal from monoclinic phases.
As can be seen in Figure~\ref{fig_fenergy}, upon decreasing temperature, \ce{K2SnX6} undergoes phase transition from cubic to tetragonal phases at 449, 433 and 281 K, while from tetragonal to monoclinic phases at 345, 301 and 210 K for X = I, Br and Cl.
From the previous experimental study of \ce{K2SnCl6}~\cite{Boysen}, the phase transition temperatures from cubic to tetragonal and from tetragonal to monoclinic phases were observed to be 262 and 255 K respectively, which are slightly lower than our predictive values of 281 and 210 K.
It should be noted that such deviations might stem from ignoring the volume change and the contributions of anharmonic modes to the phonon DOS in the calculation of Helmholtz free energies.

\section{Conclusions}
In conclusion, we have performed first-principles calculations to predict the structural, electronic, optical properties and phase stabilities of vacancy-ordered double perovskites \ce{K2SnX6} (X = I, Br, Cl) in the cubic, tetragonal, and monoclinic phases.
Our calculations reveal that the energy band gaps, effective masses of electrons and holes, and exciton binding energies increase as the symmetry of phase lowers for a given compound and as the size of halogen anion increase for a given phase.
In particular, the band gaps and exciton binding energies of the cubic \ce{K2SnBr6} and monoclinic \ce{K2SnI6} were calculated to be 1.65 eV and 59.4 meV for the former case and 1.16 eV and 15.3 meV for the latter case, providing a conclusion that on account of the suitable band gaps and optical properties the cubic \ce{K2SnBr6} and monoclinic \ce{K2SnI6} are the promising candidates for light-absorber material in PSCs.
Through the phonon calculations, the cubic and tetragonal phases were found to present the anharmonic phonon modes, whereas the monoclinic phases did not, being these anharmonic features attributed to vibrations of K and X atoms identified by the atomic resolved phonon DOS.
Finally, we calculated the Helmholtz free energy differences of the cubic from the tetragonal and of the tetragonal from the monoclinic phases, giving the phase transition temperatures as 449, 433 and 281 K for cubic-tetragonal transition, and 345, 301 and 210 K for tetragonal-monoclinic transition for X = I, Br and Cl.
Our calculations provide a comprehensive understanding of material properties of vacancy-ordered double perovskites \ce{K2SnX3}, being helpful to devise low cost and high performance PSCs.

\section{Computational Methods}
All calculations were performed by using the norm-conserving pseudopotential (NCPP) plane wave method as implemented in the ABINIT package~\cite{abinit09}.
We generated the Troullier-Martins type soft NCPPs by using the FHI98PP code~\cite{Fuchs} to describe the interaction between the ions and valence electrons, using the valence electronic configurations of the atoms, Cs--5s$^2$5p$^6$6s$^1$, Cl--3s$^2$3p$^5$, Br--4s$^2$4p$^5$, I--5s$^2$5p$^5$, and Sn--5s$^2$5p$^2$.
We employed the Perdew-Burke-Ernzerhof functional (PBE)~\cite{PBE} within generalized gradient approximation (GGA) for the exchange-correlation interactions between the valence electrons.
The cutoff energy for plane wave basis sets and Monkhorst-Pack $k$-points for electron density were set to be 40 Ha and (6$\times$6$\times$6) for cubic and (6$\times$6$\times$4) for tetragonal and monoclinic phases, giving a total energy accuracy of 5 meV per formula unit.
The variable-cell structural optimization were performed until the forces acting on atoms were less than 10$^{-5}$ Ha/Bohr with a tight self-consistent convergence threshold of 10$^{-14}$ Ha for total energy.

In order to obtain the reliable description on the electronic structures, we calculated the electronic band structures by using the PBE and the Heyd-Scuseria-Ernzerhof (HSE06) hybrid functionals~\cite{HSE04jcp} with and without the spin-orbit coupling (SOC) effect.
We replaced 20~\% of the PBE exchange functional by the exact Hartree-Fock exchange functional, producing the energy band gaps in good agreement with experiment for halide perovskites~\cite{jong19ic, HSEDu}, and considered the SOC effect only when calculating the electronic structures.
The optoelectronic properties including frequency-dependent dielectric functions, light-absorption coefficients, effective masses of electron and hole, and exciton binding energies were estimated by using the computational methods detailed in our previous works~\cite{Jong16prb, jong2017jps, jong19ic}.

To assess the phase stability of \ce{K2SnX6}, we calculated the phonon dispersions and phonon DOS by using the DFPT method as implemented in ABINIT package with a tighter convergence threshold of 10$^{-18}$ for potential residual.
By post-processing the calculated phonon DOS, we evaluated the constant-volume Helmholtz free energies of the cubic, tetragonal and monoclinic phases for \ce{K2SnX6} with increasing the temperature from 0 to 1000 K with an interval of 10 K.
To obtain more reliable phonon DOS and Helmholtz free energies, finer meshes of (80$\times$80$\times$80) for cubic and (100$\times$100$\times$80) for tetragonal and monoclinic phases were used.
With the calculated Helmholtz free energies, we calculated the temperature of phase transition from one phase to another phase by free energy differences.

\section*{Acknowledgments}
This work is supported as part of the fundamental research project ``Design of Innovative Functional Materials for Energy and Environmental Application'' (No. 2016-20) funded by the State Committee of Science and Technology, DPR Korea. Computation was done on the HP Blade System C7000 (HP BL460c) that is owned by Faculty of Materials Science, Kim Il Sung University.

\bibliographystyle{elsarticle-num-names}
\bibliography{Reference}

\end{document}